# CIDR: A Large-Scale Industrial Source Code Dataset for Software Engineering Research

Vladislav Savenkov[1]

[1]*Fermatix AI*

*hi@fermatix.ai*

## Abstract

We present the Curated Industrial Developer Repository (CIDR), a large-scale dataset of real-world software repositories collected from industrial partners. The dataset comprises 4,687 repositories spanning 84 programming languages, totaling 1.03 billion raw lines of code (670 million logical lines), along with structured metadata at the repository level, full version control history, and engineering-practice attributes such as continuous integration usage and the presence of automated tests. All repositories were collected, filtered, and anonymized through a multi-stage pipeline developed specifically for this purpose. We additionally report an exploratory fine-tuning study that adapts a 3-billion-parameter code language model to CIDR and quantifies the effect on held-out enterprise code. CIDR is intended to support research in code intelligence, software quality analysis, developer tooling, and related software engineering tasks. Access to CIDR is provided under a restricted license; details on eligibility and terms are available at https://fermatix.ai/#Contact.

## 1. Introduction

The availability of large, high-quality source code datasets is a prerequisite for progress in code intelligence and software engineering (SE) research. Models for code completion, bug detection, code summarization, and other downstream tasks are only as reliable as the data they are trained on. While several public dataset efforts exist, they predominantly draw from open-source platforms such as GitHub, which may not capture the diversity and complexity of industrial software.

We address this gap by introducing CIDR — a dataset built in close collaboration with industrial partners who directly contributed their proprietary repositories. This arrangement affords access to software that is typically unavailable to the research community, while our anonymization and vetting pipeline ensures that sensitive information is protected and data quality is maintained.

The principal contributions of this paper are as follows:

- A large-scale, industry-sourced repository dataset with full version control history.
- A reproducible, multi-stage data collection and anonymization pipeline.
- A structured metadata schema and an associated data management system.
- A detailed characterization and quality analysis of the resulting dataset.

# 2. Related Work

## 2.1 Existing Code Datasets

The availability of large-scale source code datasets has been a key driver of progress in code intelligence over the past decade. CodeSearchNet (Husain et al., 2019) is one of the most widely used benchmarks in the field, comprising approximately six million functions scraped from open-source repositories on GitHub, spanning six programming languages: Go, Java, JavaScript, PHP, Python, and Ruby. The corpus was designed primarily to support research on semantic code search, and pairs function bodies with their corresponding docstrings as natural language annotations.

More recently, the BigCode project (Kocetkov et al., 2022) released The Stack, a dataset of permissively licensed source code intended to support the pre-training of large language models for code. The initial release contained approximately 3.1 TB of source code covering 30 programming languages, subsequently expanded to over 6 TB across 358 languages in the updated version. A central design goal of The Stack was transparency and data governance: the authors restrict the corpus to permissively licensed code and provide a public opt-out mechanism allowing developers to request removal of their contributions from the dataset. In a similar vein, CodeParrot (Tunstall et al., 2022) introduced a 180 GB Python-centric dataset derived from a Google BigQuery dump of GitHub, along with a deduplicated variant, demonstrating that near-deduplication is critical for achieving competitive performance on downstream code generation benchmarks.

Beyond these general-purpose datasets, a number of domain-specific or task-specific collections have been proposed. PY150 (Raychev et al., 2016) focuses on Python abstract syntax trees for program synthesis tasks, while CONCODE (Iyer et al., 2018) targets Java code generation from natural language descriptions. CodeXGLUE (Lu et al., 2021) assembles a collection of downstream benchmarks across multiple code intelligence tasks, drawing on several of the corpora mentioned above. Despite their diversity, all of these datasets share a defining characteristic: the source material is drawn exclusively from publicly accessible, open-source repositories hosted on platforms such as GitHub or GitLab.

## 2.2 Limitations of Open-Source-Only Datasets

The exclusive reliance on open-source repositories introduces several systematic biases that limit the representativeness of existing code datasets.

First, there is a strong license bias. Datasets that restrict themselves to permissively licensed code — such as MIT, Apache 2.0, or BSD — exclude a large portion of publicly available repositories governed by copyleft licenses, and exclude proprietary software entirely. As a result, the code that models are trained on may reflect the conventions and idioms of a particular subset of the developer community, rather than the broader software industry.

Second, open-source repositories tend to underrepresent the scale and complexity of enterprise software. Industrial codebases commonly span hundreds or thousands of interconnected modules, exhibit long development histories measured in decades, and encode domain-specific business logic that does not appear in hobby or academic projects. The architecture patterns, dependency management practices, and coding standards prevalent in large organizations are therefore poorly represented in existing datasets.

Third, the language and ecosystem coverage of publicly mined datasets is skewed towards languages with strong open-source communities, most notably Python, JavaScript, and PHP. Languages prevalent in industrial and embedded systems contexts — such as Lisp, AppleScript, Arduino, or various domain-specific languages — are substantially underrepresented or absent from existing collections.

Fourth, the absence of structured metadata in most open-source repositories limits the downstream utility of the datasets. Commit messages are often terse or uninformative, issue trackers may not be publicly accessible, and the domain or business function of the software is rarely documented. This reduces the feasibility of tasks such as domain-conditioned generation, requirement-to-code mapping, or automated code review that would benefit from richer contextual information.

## 2.3 Industrial Collaboration in SE Research

The challenge of obtaining access to high-quality, industry-grade software artifacts has long been recognized in the empirical software engineering community. Briand (2012) noted that a fundamental barrier to industrial relevance in SE research is the limited availability of real-world systems for empirical study, and advocated for context-driven research grounded in the concrete needs of practitioners. More recently, several systematic studies have examined the structure and outcomes of industry-academia collaboration (IAC) projects in software engineering. Garousi et al. (2019) analyzed 101 IAC projects across 21 countries and found that, while such collaborations produce measurable innovation benefits, their prevalence remains lower than the community would desire, partly due to difficulties in data sharing and misaligned timelines between research and commercial cycles. Marijan and Sen (2022) characterized patterns and anti-patterns of knowledge co-creation in such collaborations, identifying artifact sharing — including datasets, code, and tools — as an area where standardized protocols are particularly valuable.

In the code intelligence and machine learning communities, a small number of prior works have obtained proprietary or semi-proprietary code through direct partnerships. Commercial tools such as GitHub Copilot and Amazon CodeWhisperer are known to have been trained on data that includes non-public code, though the composition of their training corpora has not been disclosed.

Within the open research literature, efforts to include industrial software have been more limited. Some SE studies (e.g., defect prediction and code smell research) have used anonymized snapshots of industrial systems obtained through bilateral agreements, but these datasets are typically not released publicly and cover only a small number of systems.

CIDR addresses these gaps by establishing a repeatable, production-grade pipeline for industrial data collection and anonymization, enabling a much larger number of contributing organizations than prior work, and making the resulting dataset available to qualified researchers under a restricted license. To our knowledge, CIDR is the first large-scale code dataset constructed through systematic, direct collaboration with industrial partners and made available to the research community under a formal access agreement.

# 3. Dataset Overview

Table 1 summarizes the key characteristics of CIDR; Section 9 provides the full statistical characterization, including engineering-practice attributes. Throughout this paper, *proprietary* denotes codebases developed as closed-source software within the contributing organizations; they have not been published on public hosting platforms, and their closed-source development within the contributing organizations, contractually represented under Section 8.1, makes prior inclusion in public training corpora unlikely.

| **Characteristic** | **Value** |
|---|---|
| Total repositories | 4,687 |
| Total commits | 3,164,663 |
| Total pull requests | 441,075 |
| Total lines of code (raw) | 1,033,810,182 |
| Total lines of code (logical) | 670,281,622 |
| Auto-generated lines of code | 189,032,349 (18.3% of raw) |
| Total characters | 45,430,661,433 |
| Total source files | 5,505,495 |
| Programming languages | 84 |
| Date range (commits) | 2012–2026 |

*Table 1. Summary statistics of CIDR. Line counts follow the definitions given in Section 4.3.2: raw lines count all physical lines including blank lines and comments; logical lines count code lines only, excluding blank lines, comments, and dependency directories; auto-generated lines are detected heuristically and reported separately. Derived from the earlier git-history extraction; commit timestamps are not recorded in the expanded catalog.*

## 4. Data Collection Pipeline

Figure 1 provides a high-level overview of the collection pipeline, which proceeds through five stages: partner acquisition, repository submission, metadata collection, repository selection, and anonymization.

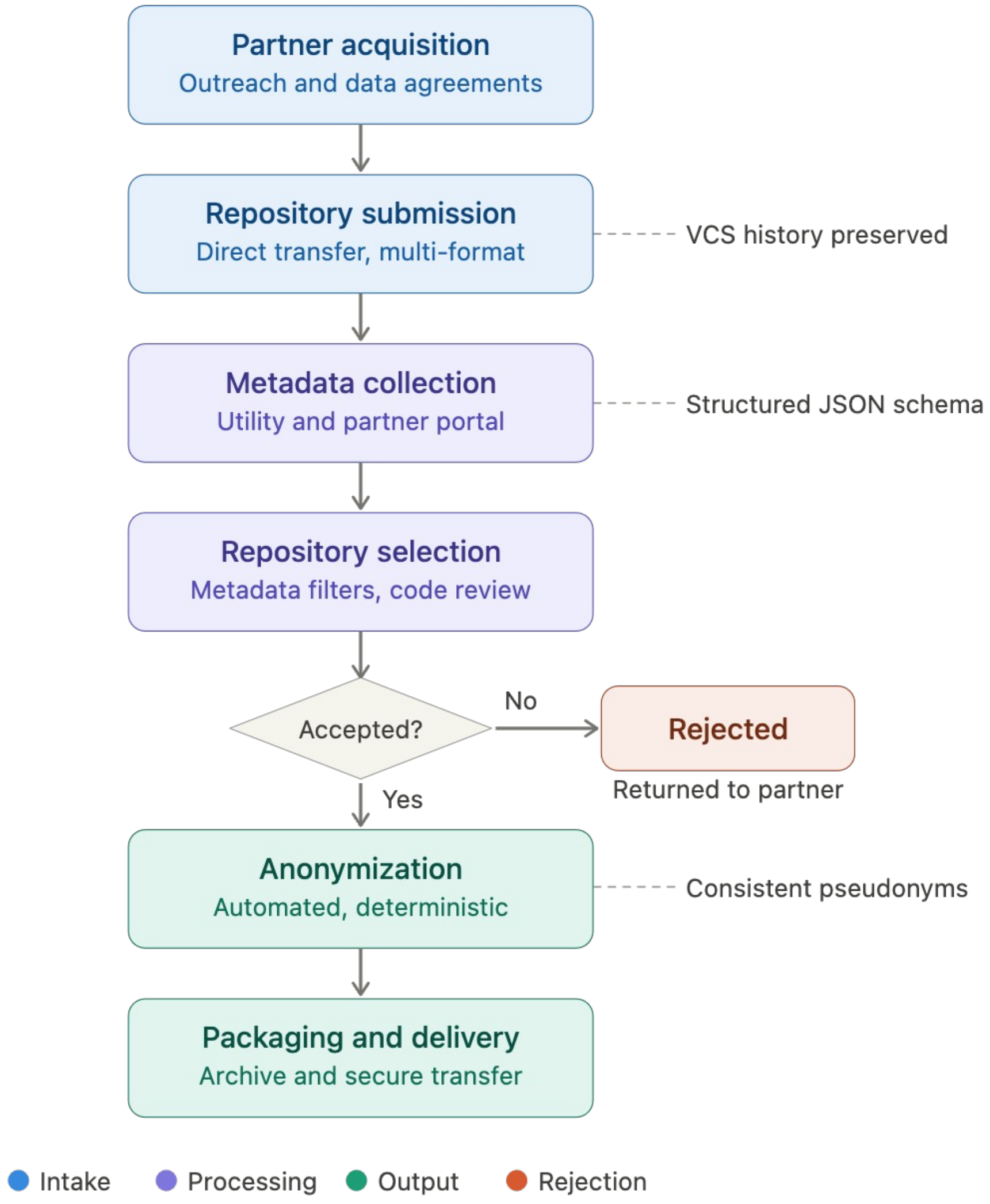

*Figure 1. Pipeline diagram.*

### 4.1 Partner Acquisition

Participating organizations were identified and recruited through a combination of outreach channels, including direct personal referrals within professional networks, targeted PR and marketing communications, presentations at industry conferences and developer events, and bilateral agreements with technology companies and software consultancies.

Initial contact was followed by a structured onboarding sequence before any data transfer took place. Each candidate organization first participated in a scoping conversation with the curation team to establish whether the nature of their software development activity matched the target profile of the dataset. Only after a positive initial assessment did the process advance to the compliance and legal stage: candidates completed a KYC questionnaire and provided the documentation required under the data sharing and anonymization agreement described in Section 8, including representations regarding intellectual property ownership, contributor rights assignment, and the human authorship of contributed code. Repository-level selection, governed by the criteria defined in Section 5, was applied subsequently — either at the pre-agreement metadata stage or following full submission, depending on the partner's preferred onboarding path. The formal agreement was executed only after both organisational and repository-level eligibility had been confirmed. Source code transfer was initiated exclusively upon agreement execution.

The primary criterion for partner eligibility was the nature and scale of the software development activity: we sought organizations whose engineering teams actively develop and maintain production codebases representative of enterprise software — that is, companies and individual development teams for whom AI-powered code assistance tools are a realistic and near-term adoption scenario. This focus ensures that CIDR reflects the kinds of codebases these tools will actually be applied to, rather than academic or hobbyist software.

In addition to domain fit, prospective partners were evaluated on their capacity and willingness to provide repositories with sufficient development history, the diversity of programming languages and application domains they could contribute, and their legal standing to enter into a data sharing and sublicensing agreement. Organizations that could not satisfy these criteria, or whose repositories did not meet the minimum quality thresholds defined in Section 5, were not onboarded.

Contributing organizations were recruited in an initial collection campaign and through continued onboarding thereafter. The outreach drew primarily on Fermatix AI's background as a software development and IT services contractor: the initial candidate pool consisted largely of companies with which the team had prior working relationships, as existing professional trust substantially reduced the friction associated with compliance negotiations and legal agreement execution. Beyond relationship-based referrals, candidates were also identified through targeted outreach at industry events and through direct approaches to companies encountered in the course of other commercial engagements.

Candidate organizations were evaluated against a set of criteria designed to ensure both legal viability and dataset diversity. The first requirement was organizational continuity: only active, operating companies were considered, as data licensing rights can only be formally transferred by a legal entity that continues to exist at the time of agreement. The second requirement concerned portfolio scope and scale — the size of the organization's engineering function and the breadth of its development portfolio were assessed to determine whether the contribution could meaningfully diversify the dataset's language and domain coverage. Organizations whose software activity was too narrow, too recent, or too small to add incremental value were deprioritized in favour of contributors with more varied and mature codebases. A further consideration, applied during the repository-level evaluation rather than at the partner selection stage, was the degree to which submitted code exhibited signs of AI-assisted generation: repositories where AI tooling had displaced substantive human authorship were excluded under the criteria described in Section 5.3, which in some cases affected the effective contribution of otherwise eligible partners.

The contributing organizations span a range of application domains — including enterprise web and mobile application development, fintech, e-commerce platforms, and custom software consultancy — and are headquartered across multiple jurisdictions. Collectively, their commit histories extend from 2012 to 2026, representing 14 years of cumulative industrial software development activity.

## 4.2 Repository Submission

Repository transfer was initiated only after the data sharing agreement had been executed by both parties. Partners submitted repositories through one of three channels, depending on their technical infrastructure and contractual preferences.

The most common submission path was a Git bundle archive. Partners were provided with a shell script (see Appendix E) that performs a complete bare-mirror clone of a repository and packages it into a .bundle file using git bundle create --all. This approach captures all refs, branches, and tags in a single portable file and requires no persistent network access after the initial clone, making it well-suited for partners operating under strict firewall or data governance policies. The script normalises repository names into a GitLab-safe slug format to prevent naming collisions during ingestion.

For codebases that had no associated version control history — typically legacy software maintained as directory archives — partners submitted a plain compressed archive of the source tree. In this case, our ingest pipeline initialised a fresh Git repository in the unpacked directory, staged all files, and created a single synthetic initial commit before producing a bundle. Repositories ingested via this path are distinguishable in the metadata by a commit count of one and a single branch, which consumers of the dataset can use to filter or stratify analyses that depend on development history. Large binary files exceeding 95 MB in these synthetic repositories were migrated to Git LFS prior to ingestion to prevent bundle size inflation.

The third channel was direct remote access. Partners granted read access to specific repositories or project groups hosted on GitHub, GitLab, or Bitbucket. Our pipeline fetched all advertised refs — including merge request and pull request refs where available — into a local bare mirror, resolved the symbolic HEAD to the remote default branch, and produced a bundle from the resulting mirror. This path was preferred for partners contributing multiple repositories simultaneously, as it avoided the overhead of manual bundle creation on their side.

Regardless of submission channel, all ingested bundles were subsequently cloned to our internal GitLab instance. Preserving full version control history — including commit authorship, timestamps, and branch topology — is essential for downstream research tasks such as code evolution analysis, defect prediction, and developer behaviour studies.

A fourth, lightweight path was available for partners unwilling to transfer source code prior to agreement execution. In this case, the partner ran the metadata utility described in Section 4.3.1 locally on their own infrastructure and transmitted only the resulting JSON metadata record. This allowed our curation team to evaluate the repository against the selection criteria defined in Section 5 — confirming language composition, commit count, contributor count, and codebase size — without receiving the source code itself. Repositories approved at the metadata stage were subsequently submitted through one of the three full channels described above once the agreement was finalised. This path was particularly important for organisations that required internal IP review or legal approval before any external data transfer, as it permitted repository-level selection to proceed in parallel with agreement negotiation.

## 4.3 Metadata Collection

Metadata were collected through two complementary mechanisms that are deliberately symmetric in their design. For repositories submitted directly, metadata were extracted automatically by the curation pipeline using the internal metadata utility described below. For repositories submitted via the partner portal, partners ran the same utility locally on their own infrastructure and uploaded the resulting record. This symmetry ensures that the metadata schema and extraction methodology are identical regardless of the submission path, eliminating any systematic differences in the derived statistics attributable to the collection mechanism itself.

### 4.3.1 Metadata Utility

The metadata utility is an open-source command-line tool available at https://github.com/Fermatix/repo_metadata_cli. Given a local clone of a repository or a bundle file, it produces a single-row CSV record capturing the fields described in Table 2. The utility combines four underlying tools: git log and git shortlog for version control history, cloc for language-aware line counting, and Tree-sitter grammars for function-level structural analysis. Where Tree-sitter grammars are unavailable for a given language, function-level metrics default to zero and are recorded as such.

The fields are organised into five categories. *Identity fields* (repo_id, repo_name) uniquely identify each repository record within the dataset; repo_id is a randomly generated UUID assigned at extraction time and serves as the primary key in all downstream metadata joins. *Language and stack fields* (languages, extensions, stack) capture the distribution of programming languages as a share of lines of code, using the language definitions from cloc; stack additionally provides a human-readable summary of the top three languages by share. *Licensing* is captured as a single enumerated field (license_type) reflecting the root-level licence detected in the repository; repositories for which no recognised licence file is present are recorded as UNKNOWN. This field has since been removed from the released metadata; see Section 8.3. *Git history fields* (created_at, commit_count, branch_count, contributors_count) characterise the development activity and team structure of the repository. *Code metrics* (files, loc, raw_loc, avg_func_length, docstring_ratio, duplication_ratio, documentation_cnt) quantify the size, structure, and documentation density of the codebase. The distinction between loc and raw_loc is noteworthy: loc counts only lines belonging to the programming languages explicitly supported by CIDR, while raw_loc counts all non-blank, non-empty lines irrespective of language — the ratio of the two provides a proxy for the proportion of the codebase that falls within scope.

Three *size fields* (repo_git_history_mb, repo_bundle_mb, repo_worktree_mb) capture storage footprint at different levels of the repository structure. These fields are not research-facing metadata in the primary sense; they are used internally during the selection and packaging stages to flag repositories with anomalously large binary assets or disproportionate history sizes relative to codebase size.

| Field | Type | Description |
|---|---|---|
| repo_id | UUID | Randomly assigned primary key |
| repo_name | string | Derived from bundle filename stem |
| languages | JSON | Language distribution by share of LoC |
| extensions | JSON | File-extension distribution by share of LoC |
| stack | string | Top-3 languages with percentages (human-readable) |
| license_type | enum | Detected root-level license (retired; see Section 8.3) |
| created_at | timestamp | First commit timestamp |
| commit_count | integer | Commits on the most recently active branch |
| branch_count | integer | Local and remote branch count |
| contributors_count | integer | Unique authors across full history |
| repo_git_history_mb | float | Size of .git directory (MB) |
| repo_bundle_mb | float | Size of bundle file (MB) |
| repo_worktree_mb | float | Size of working tree excluding .git (MB) |
| files | integer | Total tracked file count |
| loc | integer | Code + comment lines (supported languages only) |
| raw_loc | integer | Code + comment lines (all languages) |
| avg_func_length | float | Average function length in lines (Tree-sitter) |
| docstring_ratio | float | Comment lines / code lines |
| duplication_ratio | float | Estimated duplication: 1 − unique_lines / total_lines |
| documentation_cnt | integer | Total line count across root-level README files |

*Table 2. Metadata fields collected per repository by the Fermatix metadata utility.*

### 4.3.2 Catalog Metrics: Provenance and Definitions

The statistics reported in Section 9 are produced by version 2 of the repository metadata utility introduced in Section 4.3.1 (publicly available at https://github.com/Fermatix/repo_metadata_cli), orchestrated by the internal delivery pipeline; Appendix H enumerates the exact tools, invocations, and rule sets per metric. Version 2 replaces the line-counting backend (scc in place of cloc), adds engineering-practice detectors, and revises several definitions; figures produced by the two versions are therefore not directly comparable, and the paragraph "Changes relative to the earlier characterization" below makes the differences explicit. All catalog metrics are computed by rule-based analysis of the repository content, its history, and — for pull requests — hosting-platform metadata; none are manually annotated. A small number of administrative fields (project status, project category, update availability) are supplied by the contributing partner and are populated on a rolling basis; a blank value in those fields means the information has not yet been provided, not that the property is absent.

**Line counting.** Three line counts are reported per repository, all computed with scc. *Raw lines of code (LoC)* is the total number of physical lines across all source files, including blank lines and comments. *Logical LoC* is the number of code lines — excluding blank lines and comments — counted outside dependency and build directories (node_modules, vendor, dist, build, bower_components); it is the filtered, language-aware measure of original source code, and it is the basis on which per-line licensing terms are defined. *Auto-generated LoC* counts code lines in files identified as machine-generated by signals including residence in a generated-code directory, generated-file naming patterns, and generator markers within the first five lines of the file (the full rule set is enumerated in Appendix H). Auto-generated lines are reported separately and never exceed the logical count by construction, as the scan is restricted to the logical-LoC file set. By construction, logical ≤ raw for every repository; the auto-generated share is 18.3% of raw lines overall.

**Language composition.** Per-repository language shares are fractions of code lines attributed to each language by scc's language taxonomy, computed outside dependency directories, with languages below a 1% within-repository share omitted. The *primary language* of a repository is the language with the largest share. The dataset-level distribution (Section 9.1) weights per-repository shares by raw LoC. In the current catalog release, 105 distinct taxonomy entries occur, of which 84 are programming languages; the remaining 21 are data, markup, styling, and documentation formats (CloudFormation (JSON), CSS, CSV, Docker ignore, HTML, INI, JSON, JSON5, LaTeX, LESS, License, Markdown, MSBuild, Plain Text, Qt Translation Source, RAML, Sass, SVG, TOML, XML, and YAML), which are not counted as programming languages.

**Duplication rate.** The fraction of duplicated code detected by jscpd, a token-based copy-paste detector, over the repository's source files, excluding dependency directories and generated, minified, and lock-file content (the full ignore list is given in Appendix H); values lie in [0, 1] (observed median 0.012, mean 0.071).

**Documentation density.** The fraction of functions and methods that carry a documentation comment, measured by Tree-sitter parsing (observed median 0.036). This differs from the comment-to-code ratio reported by the version-1 utility.

**Average function length.** The mean function length in lines — measured over the full definition, signature through closing delimiter — across all functions parsed by Tree-sitter (observed median 11.5 lines; distribution in Figure A.4).

**Version-control and collaboration metrics.** *Commit volume* counts non-merge, non-revert commits on the repository's default branch. *Total PRs* counts merged pull or merge requests, retrieved from the hosting platform's API where available and otherwise detected from merge patterns in the git history. *CI on PRs* records whether continuous-integration configuration is present in the repository (for example, workflow directories or pipeline files).

**Engineering-practice attributes.** The ordinal attributes are assigned by rule-based detectors. *Deployment tier*: Enterprise if infrastructure-as-code or Kubernetes artifacts are present (Terraform files, Helm charts, deployment manifests); otherwise Full CI-CD if the CI configuration contains deployment keywords (deploy, release, publish, ship); otherwise Basic CI if CI configuration exists; otherwise None. *Observability level*: Full SRE if distributed-tracing instrumentation (OpenTelemetry, Jaeger, Honeycomb) is referenced in source files; APM+Alerting if application-performance-monitoring instrumentation is found; Basic if logger instantiation or console-logging calls are present; otherwise None. Detection is restricted to source files, excluding vendored code and CI configuration, so that tooling that is referenced but not used is not counted. *Test coverage*: Comprehensive if at least ten test files or three distinct test directories are found (compiled-language test files are additionally verified by framework markers in their content, not filename alone); Basic if any test files or a test-framework configuration are present; otherwise None. *README quality*: Comprehensive if a documentation directory or contribution guide accompanies a README covering setup and usage; Detailed if the README covers setup together with usage or architecture; Basic if a README exceeds 200 characters; otherwise None. *Containerization* records the presence of container or orchestration configuration (for example Dockerfile and Compose files; the full trigger set is given in Appendix H). *Issue-tracking linkage*: Linked to Commits if issue references occur in recent commit messages (upgraded to Full+Design Docs if the repository also carries RFC or architecture-decision-record directories); Basic if issue templates are present; otherwise None.

**Characters and file counts.** *Characters* is the total Unicode character count over the same file set as logical LoC; it is the quantity most directly proportional to language-model token counts. *Source file count* is the number of source files recognized by the counter.

**Changes relative to the earlier characterization.** Three headline figures changed definition between the utility versions. (1) *Lines of code*: the version-1 loc counted code and comment lines in supported languages (373 million over the earlier 2,545-repository snapshot); the catalog reports raw (1.03 billion) and logical (670 million) lines over the expanded collection under the definitions above. The version-1 field raw_loc referenced in Sections 4.3.1, 5.1, and 9.4 counted non-blank lines irrespective of language and is likewise distinct from the raw LoC defined here. (2) *Languages*: 138 under the version-1 cloc taxonomy, which enumerated markup, configuration, and data formats and applied no within-repository threshold, versus 105 distinct entries under the version-2 taxonomy with the 1% threshold, of which 84 are programming languages (75 in the July 2026 catalog release on which Section 9 is based); the reduction reflects the changed definition, not removed content. (3) *Duplication*: the version-1 ratio was defined as 1 − unique lines / total lines within a repository (median 0.58); the version-2 rate is block-level jscpd detection (median 0.012). The two measure different phenomena — line-level repetition including boilerplate versus copy-paste blocks — and are not comparable; no conclusion about a change in dataset redundancy should be drawn from the difference.

## 5. Repository Selection and Quality Criteria

Each submitted repository undergoes a two-stage selection process: automated filtering by metadata, followed by manual review of the source code. The two stages are complementary rather than redundant — automated filtering eliminates structurally unsuitable repositories at low cost, while manual review applies qualitative judgements that cannot be reduced to simple thresholds.

### 5.1 Metadata-Based Filtering

The automated stage applies two hard requirements derived from the metadata record produced by the utility described in Section 4.3.1. First, the repository must contain at least 1,000 lines of code as measured by the loc field — that is, lines belonging to programming languages explicitly supported by CIDR, excluding markup, configuration, and other non-code content captured by raw_loc. This threshold excludes stubs, single-file snippets, and repositories that are nominally versioned but contain negligible original code. Second, the repository must be non-empty and fully parseable: the metadata utility must complete without error, and the bundle must yield a valid working tree. Repositories that fail either condition are rejected without appeal. The sole exception is a technical parsing failure — that is, a case where the metadata utility terminates with an error due to a malformed bundle or an environment issue unrelated to the repository's actual content — in which case the partner may resubmit a corrected archive.

### 5.2 Manual Code Review

Repositories passing the automated stage are reviewed manually by members of the curation team. Each repository is assessed by at least one reviewer, with a subset reviewed independently by two reviewers to calibrate inter-reviewer consistency and guard against individual blind spots. Reviewers examine two complementary objects: the metadata record and the source code itself. The metadata review checks for internal consistency — for example, whether the declared language distribution is plausible given the file count and loc value, and whether the commit history reflects genuine development activity rather than a bulk import. The code review assesses whether the repository constitutes authentic, production-grade software: reviewers inspect representative source files, directory structure, commit messages, and architectural organisation to form a holistic judgement of software quality and development maturity.

### 5.3 Rejection Criteria

Rejection may occur at either stage. Table 3 summarizes the criteria applied at each stage; the two criteria marked with an asterisk are discussed in further detail below.

| Stage | Rejection criterion |
|---|---|
| Metadata | loc value below the 1,000-line threshold |
| Metadata | Metadata record unparseable or internally inconsistent |
| Code review | Only auto-generated, scaffolded, or third-party code with no original contribution |
| Code review | Fork or near-duplicate of a repository already present in the dataset |
| Code review | Not a software project (data archive, configuration dump, static asset collection) |
| Code review | Partial codebase lacking structural completeness for project-level analysis |
| Code review | Interview assignment, coursework submission, or tutorial repository |
| Code review | Substantial evidence of AI-assisted generation without meaningful human authorship* |
| Code review | Code quality or architectural design below mid-level professional standard* |

*Table 3. Rejection criteria applied at the metadata and code review stages.*

Two criteria warrant particular elaboration. Repositories are rejected when reviewers identify substantial evidence of AI-assisted generation without meaningful human authorship — that is, cases where the codebase appears to have been produced by prompting a code generation model with minimal review, resulting in structurally shallow, semantically incoherent, or stylistically uniform code that does not reflect genuine engineering decisions. The inclusion of such repositories would introduce a qualitatively different signal into the dataset and undermine its value as a representation of human software development practice.

Repositories are also rejected when the overall code quality or architectural design falls below the standard expected of a professional mid-level software developer. This criterion is intentionally qualitative: reviewers assess factors such as separation of concerns, consistent naming conventions, error handling practices, and the appropriateness of abstractions relative to the problem domain. The rationale is that CIDR is intended to reflect the kinds of codebases that AI-powered developer tools will encounter in realistic enterprise deployments, and repositories of systematically poor quality would skew the dataset toward artefacts that do not represent that target context.

## 6. Anonymization

Anonymization was applied to every accepted repository prior to release using repo-sanitizer, an open-source command-line utility developed specifically for this pipeline and available at

https://github.com/Fermatix/repo-sanitizer. The tool operates on the complete repository — including the working tree at the time of submission and the full version control history across all branches and tags — producing a self-contained Git bundle in which potentially identifying information has been detected and replaced. Although partners may run the utility independently on their own infrastructure prior to submission, anonymization was in practice performed by the curation team on our side as part of the standard ingestion workflow.

## 6.1 What Is Anonymized

The utility targets five categories of potentially identifying information:

| Category | Examples |
|---|---|
| Commit author identities | Names, email addresses, committer fields across all branches |
| Credentials and secrets | API keys, access tokens, passwords, bearer tokens |
| Personal and organisational names | Person names and organisation identifiers in comments, docstrings, and commit messages |
| Internal network infrastructure | Private IP addresses (RFC 1918), internal hostnames, organisation-specific domain names |
| Partner-specific terms | Project codenames, client names, internal identifiers supplied via partner dictionary |

## 6.2 Anonymization Process

The pipeline executes in ten sequential steps, grouped into three phases.

*Phase 1 — Working tree.* The working tree is scanned for sensitive content using five complementary detectors. Identified entities are replaced, and a post-scan verifies that no findings remain.

*Phase 2 — Repository history.* The history is processed in two passes. Commit metadata (author names, email addresses, commit messages) is scanned and subsequently rewritten. File content is scanned at the blob level across all unique objects in the history, covering all branches and tags, followed by a full rewrite via git-filter-repo.

*Phase 3 — Gate check.* A hard pass/fail gate is enforced at the close of the pipeline. Any remaining secret, high-severity PII finding, forbidden file, or unredacted internal endpoint causes the pipeline to exit with a non-zero code, and the repository is withheld from the dataset pending manual investigation.

Detection relies on five mechanisms operating in combination:

| Detector | Mechanism | Targets |
|---|---|---|
| SecretsDetector | gitleaks subprocess | API keys, tokens, credentials |
| RegexPIIDetector | Compiled regex patterns | Emails, phone numbers, IPv4, JWTs, HTTPS URLs |
| EndpointDetector | Rule-based matching | RFC 1918 addresses, internal TLDs (.internal, .corp, .local, .lan) |
| DictionaryDetector | Aho-Corasick over partner-supplied .txt files | Codenames, client names, custom terms |
| NERDetector | Multilingual BERT transformer (Babelscape/wikineural-multilingual-ner) | Person names (PER), organisation names (ORG) (F1 ≈ 90%) |

For code files, scanning and redaction are confined to tree-sitter zones — comments and string literals — leaving identifiers, function names, and code structure entirely intact. For configuration and documentation files, the entire file content is subject to scanning. This design preserves the functional semantics of the source code while eliminating the surface area where sensitive strings most commonly appear.

All replacements are deterministic: a given original value always yields the same pseudonym for a fixed salt, computed as the first 12 hexadecimal characters of HMAC-SHA256(salt, value). This preserves structural relationships across the dataset — for example, a single author who appears across hundreds of commits is consistently replaced by the same pseudonym, maintaining commit-graph topology and contributor-count statistics without revealing the original identity. The salt is held internally by the curation team and is not included in the released dataset. Replacement masks follow a structured schema:

| **Entity type** | **Output mask** |
|---|---|
| Person name | Person_{hash12} |
| Organisation name | Org_{hash12} |
| Email address | user_{hash12}@example.com |
| Author email (commit history) | author_{hash12}@example.invalid |
| Secret (gitleaks) | REDACTED_{hash12} |
| Internal domain | {hash8}.example.invalid |
| Private IP address | 192.0.2.{1–254} |
| Regex pattern match | [name:{hash12}] |

## 6.3 Limitations

Several residual risks remain after automated anonymization.

*History blob coverage.* The NER model and the secrets detector are not applied to file content at the blob-scanning stage. In Git, a blob is the object type used to store the raw contents of a file at a specific point in history; each unique file version is stored as a separate blob in the repository's object store, meaning that a long-lived repository may contain thousands of blob objects representing every revision of every file across all branches and tags. Invoking per-blob ML inference and subprocess-based secret detection over such a corpus is computationally prohibitive. History blob scanning instead relies on the regex, dictionary, and endpoint detectors, which provide strong coverage for structured patterns but may miss free-form names or credentials in narrative text within historical file content.

*Stylometric re-identification.* Indirect re-identification remains a theoretical risk. Distinctive coding style, idiosyncratic naming conventions, domain-specific abstractions, or non-English natural language in comments may allow an informed adversary to associate a repository with a specific organization even in the absence of explicit identifiers. No automated mitigation for stylometric re-identification is currently applied; the data sharing agreement with each contributing partner explicitly addresses this residual risk and restricts the permitted uses of the data accordingly.

*Non-English NER coverage.* Content written in languages not well represented in the NER model's training data may yield lower recall for named-entity detection. Partners contributing repositories with significant non-English comment volume were advised to supplement automated anonymization with a manual review pass.

# 7. Data Management and Accounting System

As the scale of the collection effort grew, the need for a robust data management system became apparent. This section describes the tooling developed to track repositories through their lifecycle and to facilitate structured interaction with contributing partners.

## 7.1 Evolution of Tooling

In the initial phase, repository tracking was performed in shared spreadsheets. While adequate for small volumes, this approach proved insufficient as the number of repositories grew: concurrent edits introduced inconsistencies, and the absence of referential integrity made it difficult to maintain a reliable audit trail. In particular, it became impractical to enforce uniqueness constraints across partner records, track the status of individual repositories through multiple pipeline stages, or attribute changes to specific actors. These limitations motivated the development of a purpose-built internal CRM.

## 7.2 Internal CRM

The CRM is a web application built around a relational database that tracks each repository as a record progressing through a defined lifecycle: submitted, under review, accepted, anonymized, packaged, and delivered. The system is built on Django 5.1 and Python 3.12, backed by a PostgreSQL 16 database, and deployed via Docker Compose behind an Nginx reverse proxy.

The data model is built around five principal entities. At its centre sits the repository record, which links each submission to a contributing partner and stores its associated metadata attributes in a schema-free structure rather than fixed database columns. This design allows the curation team to extend or modify the set of tracked attributes at runtime, without requiring application changes or database migrations. The schema itself is defined by a separate entity that specifies, for each attribute, its data type, validation rules, display ordering, and the visibility and editability permissions exposed to partners. A third entity maintains a registry of the legal organisations associated with each partner, capturing the signatory and jurisdictional information relevant to the data sharing agreements described in Section 8. Partner accounts are extended with a profile entity that supports hash-based authentication, enabling partners to access the system without exposing internal account identifiers in external communications. Finally, an audit log entity records every significant operation in the system — including record creation, field updates, and bulk imports — as an immutable, timestamped entry attributed to the acting user, providing a complete and tamper-evident history of all data contributions.

Role-based access is enforced at the view layer through two permission decorators. Administrator accounts hold unrestricted access to all records, the full column schema, the legal entity registry, and the audit log. Partner accounts are strictly isolated: all database queries in the partner-facing interface are filtered by the authenticated user's identity at the ORM level, and attempts to access records belonging to other partners return a 404 response rather than a 403, avoiding disclosure of record existence.

### 7.3 Partner Portal

Partners interact with the system through a dedicated self-service portal accessible via a web browser. Authentication uses a hash token issued by the curation team at onboarding, rather than a conventional username, which avoids exposing internal account identifiers in external communications.

The portal provides the following capabilities. Partners may submit new repositories by completing a guided metadata entry form, in which each field is rendered according to its declared type — dropdown selectors for enumerated attributes, date pickers for temporal fields, and validated URL inputs for repository addresses. Once submitted, partners can monitor the current lifecycle status of each repository and view its full status history. Metadata records can be reviewed and enriched after submission: fields designated as partner-editable by the curation team remain modifiable both through the full edit form and via inline cell editing directly within the table view, while read-only fields are displayed but locked. Partners may also respond to queries raised by the curation team, which are surfaced within the portal. Finally, the portal supports bulk operations through CSV import and export, allowing partners to prepare and update multiple repository records externally and upload them in a single operation, subject to the same field-level editability constraints that govern the form-based interface.

All partner actions — record creation, field modifications, and CSV imports — are captured synchronously in the audit log, providing the curation team with a complete and attributable history of every data contribution.

## 8. Ethical Considerations and Licensing

### 8.1 Partner Consent and Data Agreements

All participating organizations entered into a Master Source Code License and Sublicensing Agreement with Fermatix AI prior to submitting any repositories. Contributing partners are legal entities — software development agencies, technology companies, and studios — not individual developers; all representations and warranties are made at the entity level. The agreement governs the full lifecycle of the contributed material: the scope and baseline of the repository transfer, the line-of-code counting methodology used to determine the accepted inventory, the conditions under which the data may be processed and sublicensed to third parties, and the information security requirements that apply to storage and delivery. Each agreement is executed per repository, with a corresponding acceptance certificate signed by both parties upon delivery.

The agreement requires partners to provide representations and warranties across four areas. First, partners must warrant that they hold full legal title to the contributed code and that all intellectual property rights have been validly assigned to them by employees and contractors through written agreements. Second, partners represent that the commit history is authentic and has not been materially altered to misrepresent authorship. Third, partners must disclose all open-source and third-party components included in the repository, and confirm the absence of any license terms that would impose copyleft obligations or other restrictions inconsistent with the sublicensing rights granted. Fourth, partners explicitly warrant that the repository was predominantly created by human developers under their direction, and that no material portion was generated by a generative AI tool in a manner that would restrict or impair the licensed rights — a representation that is contractually aligned with the AI-generation rejection criterion described in Section 5.3; under the AI-tool disclosure obligations of Schedule 7, the warranty extends to third-party model terms of use, distillation or training restrictions, and other encumbrances attaching to code produced with AI assistance, and repositories whose disclosures identify such restrictions are not accepted.

The agreements also incorporate a human authorship and AI tool disclosure form (Schedule 7) and a KYC compliance questionnaire (Schedule 5), which together establish legal accountability for the authenticity and provenance of the contributed material.

## 8.2 Regulatory Compliance

The collection and processing of repository data involves incidental handling of personal data — principally commit author names and email addresses embedded in the version control history of contributed repositories. These data are processed for the sole purpose of anonymization and are not retained in any artifact included in the released dataset. Following the anonymization pipeline described in Section 6, all commit author identities are replaced with deterministic pseudonyms and the original values are recorded exclusively in the redaction manifest, which is retained internally and not distributed.

The agreements executed with contributing partners include explicit data protection provisions consistent with Regulation (EU) 2016/679 (GDPR). Each party is designated as an independent data controller with respect to any personal data processed in connection with the agreement, and each party assumes responsibility for ensuring a valid lawful basis for its processing and for providing any required notices to data subjects. Where personal data is transferred outside the European Economic Area, the agreements require that such transfer be effected pursuant to an applicable adequacy decision or appropriate safeguards, including, where required, the Standard Contractual Clauses adopted by the European Commission. Partners additionally represent that the contributed repositories do not contain personal data beyond what has been expressly disclosed and approved in writing, and accept indemnification liability for any breach of this obligation.

The anonymization step described in Section 6 serves as the primary technical measure for eliminating personal data from the dataset prior to release. The gate check enforced at the close of the pipeline provides a machine-verifiable guarantee that no high-severity PII finding remains in any delivered artifact.

## 8.3 Dataset License

CIDR is released under a commercial proprietary license. Access is granted exclusively through a formal sublicensing agreement with Fermatix AI; prospective users must submit an eligibility request through the Fermatix AI contact page at https://fermatix.ai/#Contact to initiate the licensing process.

Earlier versions of the released metadata included a license_type field recording the license file detected at the repository root; entries showing open-source identifiers were found, on inspection, to originate from framework starter templates or embedded third-party components, and UNKNOWN denoted the absence of a root-level license file. Because the field was reasonably but incorrectly read as the license of the repository as a whole, it has been removed from released metadata; the rights to each contributed codebase are governed by the data sharing agreement with the contributing partner, not by root-level license files.

The license structure reflects the deferred-royalty model defined in the partner agreements: the execution of a sublicensing agreement with a third-party user constitutes a trigger event that activates the royalty obligation owed by Fermatix AI to each contributing partner, calculated at a fixed rate per line of code as recorded in the accepted delivery inventory. This arrangement enables controlled distribution of the dataset while preserving a contractual and financial relationship between the curation team and the contributing organizations.

Redistribution of the dataset or any derivative thereof, and any use that would re-identify contributing organizations or individuals, are explicitly prohibited under the license terms. Users are required to cite the dataset in any publications that make use of it, and to report any inadvertent discovery of residual sensitive content to the curation team promptly.

# 9. Dataset Statistics and Analysis

This section provides a quantitative characterization of the repositories in CIDR. All statistics in this section and in Appendix A were computed with version 2 of the repository metadata utility (Section 4.3.2) over the July 2026 catalog release of 4,225 repositories; the current catalog, summarized in Table 1, comprises 4,687 repositories. Extended tables and figures are available in Appendix A.

## 9.1 Language Distribution

CIDR exhibits a pronounced concentration in two languages: PHP accounts for 40.5% of all raw lines of code (LoC), and JavaScript for a further 27.9%, together comprising over two-thirds of the dataset. TypeScript (5.9%) forms the third-largest share, followed by C and C++ header files (5.4% combined), C# (3.7%), and Python (3.5%). The remaining share is distributed across a further 68 languages, none exceeding 2%. In terms of repository count rather than LoC share, JavaScript is the most widely occurring language, appearing in 1,690 repositories, followed by PHP (1,004) and TypeScript (839); Python occurs in 541. The divergence between LoC dominance (PHP) and occurrence frequency (JavaScript) reflects the structural difference between large monolithic web backends and the many smaller JavaScript components that accompany them.

Measured by primary language per repository, the distribution is more diverse: PHP constitutes the primary language in 18.8% of repositories, followed by TypeScript (16.5%), JavaScript (15.7%), Python (10.9%), Swift (5.7%), Java (5.0%), Kotlin (4.6%), and Dart (3.3%). The remaining 19.3% of repositories have a primary language outside the top eight, including C# (3.1%), Shell (2.4%), and Vue (2.1%). The dominance of web and mobile development languages is consistent with the target domain of CIDR: organizations whose engineering teams are plausible near-term adopters of AI-powered developer tools. Figure A.1 and Figure A.2 provide the language distribution charts; the complete language table is given in Table A.1.

The language count depends on the counting definition. Earlier versions of this characterization enumerated 138 languages under the version-1 tool's cloc taxonomy, which included markup, configuration, and data formats (for example JSON, HTML, and CSS) and applied no prevalence threshold. The current distribution uses the version-2 taxonomy and omits languages below a 1% share within each repository (Section 4.3.2), which prunes trace occurrences; 75 languages remain. The change is one of definition rather than dataset composition.

## 9.2 Repository Size Distribution

Table 4 reports key percentiles for the principal size metrics. The distribution of lines of code is right-skewed: the median repository contains 14,157 raw lines of code (10,384 logical), while a long tail of larger codebases inflates the mean (197,024 raw) substantially. The LoC bucket breakdown in Figure A.3 illustrates this structure: the modal bucket is 20K–100K raw lines (1,021 repositories, 24.2%), and 320 repositories (7.6%) exceed 500K lines, representing large enterprise-scale systems. At the lower end, 586 repositories (13.9%) fall below 1,000 raw lines; the 1,000-line criterion of Section 5.1 was applied at collection time, and repositories below the threshold were retained where the manual review of Section 5.2 confirmed non-trivial original code.

| Metric | Median | Mean | P10 | P75 | P90 |
|---|---|---|---|---|---|
| Lines of code (raw) | 14,157 | 197,024 | 681 | 67,088 | 337,270 |
| Lines of code (logical) | 10,384 | 137,434 | 521 | 48,484 | 230,711 |
| Commit count | 46 | 378.5 | 1.4 | 232 | 809 |
| Pull requests | 1 | 74.9 | 0 | 27 | 152.6 |
| File count | 160 | 1,071.8 | 14 | 597 | 2,102.4 |
| Contributors† | 5 | 10.4 | 1 | 12 | 22 |

*Table 4. Key size percentiles across the 4,225 cataloged repositories. Full descriptive statistics including P25 and P95 are reported in Table A.2. †Contributor counts are git-derived (unique commit authors across the full history) from the earlier 2,545-repository extraction and have not been recomputed for the expanded catalog; contributor percentiles are therefore not included in Table A.2.*

The commit count distribution (median 46, mean 379) and the pull request distribution (median 1, mean 75) indicate that the catalog spans both long-lived, actively developed systems and smaller or shorter-lived projects; 25% of repositories carry more than 232 commits, and 10% more than 809. Auto-generated code, detected heuristically and reported as a separate measure, accounts for 3.3% of raw lines overall and is absent from the median repository.

Code quality metrics show a median duplication rate of 0.012 (mean 0.071), a median documentation density of 0.036, and a median average function length of 11.5 lines (Figure A.4). Definitions of these metrics are given in Section 4.3.2; the duplication rate reported here is not comparable to the duplication ratio published in the earlier characterization, which used a different measure (Section 4.3.2, "Changes relative to the earlier characterization").

## 9.3 Temporal Coverage

The current catalog does not include commit timestamps; the temporal characterization below therefore describes the previously analyzed subset of 2,545 repositories and has not been recomputed for the expanded catalog. The first-commit timestamps of accepted repositories span from 2011 to 2026, with the distribution peaking in 2025 (343 repositories) and 2021 (322 repositories). The period 2017–2021 accounts for the largest share of the dataset by repository count, reflecting the growth of the contributing partner organizations during those years. The small number of repositories with first commits after 2025 (41 in 2026) reflects repositories that were recently initialized and submitted during the final stages of the collection campaign.

The temporal distribution indicates that CIDR captures software developed across approximately 15 years of industrial practice, covering a period of substantial change in programming language ecosystems, development tooling, and software architecture patterns. Repositories with first commits before 2015 (total: 111, or 4.4% of the accepted set) tend to exhibit larger commit counts and contributor bases, consistent with longer development histories. The full temporal breakdown by year is shown in Figure A.5.

## 9.4 Submission and Acceptance Rate

The submission funnel reported in this subsection describes the initial collection campaign; repositories onboarded after that campaign undergo the same two-stage selection process described in Section 5. The current catalog comprises 4,687 repositories. During that campaign, a total of 4,449 repository submissions were received across all contributing partners. Of these, 2,545 (57.2%) were accepted following the two-stage selection process described in Section 5. A further 1,837 repositories (41.3%) were rejected at the code review stage; 66 records (1.5%) had an indeterminate or missing status, attributable to submissions that were withdrawn by partners during the evaluation process; and 1 repository (0.0%) remained under active review at the time of dataset compilation.

Rejection at the code review stage was attributable to a small number of recurring categories, as shown in Figure A.8. The dominant rejection criterion by volume was insufficient lines of code, accounting for approximately 770 rejections — repositories that cleared the automated 1,000-line threshold but were assessed by reviewers as containing negligible original code once auto-generated, third-party, or configuration content was discounted. The second most frequent category was evidence of AI-assisted generation without meaningful human authorship (approximately 40 repositories), followed by partial codebases (approximately 15), forks or near-duplicates (approximately 3), and repositories that did not constitute software projects in the conventional sense (approximately 3).

The overall acceptance rate of 57.2% reflects a deliberately selective curation policy. The rejection categories listed above, particularly the LoC-based criterion applied during review, indicate that a substantial share of submitted repositories were either small utility scripts or projects whose apparent size was inflated by vendored dependencies and generated files. This pattern reinforces the rationale for applying loc rather than raw_loc as the primary size filter, and for complementing the automated threshold with human judgment.

### 9.5 Engineering Practice Attributes

Beyond size and language composition, the catalog records categorical attributes describing the engineering practices evident in each repository; Figure A.9 summarizes their distributions. Continuous integration checks on pull requests are present in 29.5% of repositories. A deployment pipeline is identifiable in 30.7% (Basic CI 6.9%, Full CI-CD 19.2%, Enterprise 4.6%). Some form of observability tooling is present in 52.9% (basic logging 42.0%, application performance monitoring with alerting 7.7%, full site-reliability tooling 3.2%). Automated tests are absent in 73.7% of repositories, basic in 16.3%, and comprehensive in 10.0% (the attribute measures the presence and extent of test files, not executed line coverage; Section 4.3.2). Container configurations are present in 36.6%. A README is present in 50.8% of repositories (basic 35.5%, detailed 13.3%, comprehensive 2.0%), and 39.5% link commits to an issue tracker.

These distributions quantify a dimension along which industrial codebases plausibly differ from the popular open-source projects that dominate public corpora: production systems in commercial settings frequently lack the public-facing documentation and community-oriented infrastructure that characterize such projects, while nonetheless carrying substantial operational tooling. We report these attributes to support stratified sampling — for example, restricting to repositories with comprehensive test coverage for execution-based research — and as a realistic reference profile of enterprise engineering practice. The attribute taxonomy and assignment procedure are described in Section 4.3.2.

## 10. Case Study: Adapting a Code Language Model with CIDR

To complement the descriptive characterization of Sections 3–9, we report an exploratory case study that measures the effect of adapting a small code language model to CIDR. The study is designed as a controlled before/after comparison: a base model is evaluated under a fixed protocol, adapted to CIDR-derived data, and re-evaluated under the identical protocol, so that any difference is attributable to the adaptation. We state the scope of the study explicitly: it uses a single 3-billion-parameter model, a single adaptation recipe, and a compute budget of one pass over roughly 300 million tokens on a single workstation GPU. The results should therefore be read as evidence about what CIDR contributes under realistic, modest-budget domain adaptation, not as a general claim about industrial data at scale.

### 10.1 Hypotheses

The differential content of CIDR relative to public code corpora is enterprise application code, concentrated in PHP, JavaScript, and TypeScript (Section 9.1). We therefore specified two directional hypotheses in advance. H1: adaptation on CIDR improves a model's ability to model and complete previously unseen enterprise code, with the effect concentrated in the dataset's dominant languages. H2: the effect is language- and domain-specific rather than a uniform capability gain — performance on languages and task types outside CIDR's focus should not improve, and may degrade, consistent with prior findings on domain-adaptive pre-training (Gururangan et al., 2020).

## 10.2 Training Corpus

From the collection snapshot available at the time of the study, we selected 1,024 repositories and extracted their default-branch working trees. Files were filtered to source code (excluding vendored, generated, minified, binary, and data-only content), deduplicated exactly by content hash, and decontaminated by token-level 13-gram overlap against a superset of the evaluation benchmarks used in this study, additionally covering MBPP-derived task sets and Rust translations: any file sharing a 13-token window with any indexed benchmark solution was removed (907 files in total; the full audit trail is provided in Appendix F). Repositories were partitioned in advance into 969 for training and 52 for held-out evaluation (three further repositories were reserved for pipeline tests); after filtering and deduplication, 905 of the training repositories contributed at least one file. The partition is by repository, so held-out measurements reflect generalization to unseen codebases rather than memorization.

The training set comprises 258,484 documents totaling approximately 296 million tokens (byte-based estimate), drawn with a language mix that reflects CIDR's profile (PHP 44.8%, JavaScript 18.3%, TypeScript 6.7%, Python 5.1%, with the remainder across further languages) and a per-repository cap of 1.5 million tokens to prevent any single codebase from dominating. Because the study targets code completion rather than instruction following, 60% of documents were formatted as fill-in-the-middle (FIM) examples using the model's native FIM control tokens — the document is split at character level into a prefix, middle, and suffix, and the model learns to reconstruct the middle from the surrounding context — and the remaining 40% as plain left-to-right continuation. Both objectives reduce to standard next-token prediction over packed 1,024-token sequences.

## 10.3 Model, Adaptation Method, and Training

We adapt Qwen2.5-Coder-3B (base variant), an open-weights code model whose published evaluation protocol we reproduce below. Full fine-tuning of a 3-billion-parameter model exceeds the memory of the available hardware (a single 11 GB NVIDIA RTX 2080 Ti), so we use quantized low-rank adaptation (QLoRA; Dettmers et al., 2023): the base weights are quantized to the 4-bit NormalFloat (NF4) data type and kept frozen, and low-rank adaptation (LoRA; Hu et al., 2022) matrices (rank 32, scaling parameter $\alpha = 32$, dropout 0.05) are trained on all linear projections — 59.9 million trainable parameters, 1.9% of the model. Training ran for 18,127 optimizer steps at an effective batch size of 16,384 tokens per step — 297 million training tokens, which the trainer measured as 1.06 epochs over the packed corpus (implying a tokenized size of approximately 280 million tokens against the byte-based estimate above) — with a peak learning rate of $2\times10^{-4}$ under a cosine schedule with 3% warm-up. Training required approximately 93 hours of compute at the measured throughput of 895 tokens per second.

During training, loss on a sample of the held-out repositories was evaluated every 500 steps (Figure 2). Held-out loss decreased steadily from 0.908 to a minimum of 0.862 at step 17,000 (equivalently, perplexity — the exponentiated loss — improved from 2.479 to 2.369) and rose slightly thereafter; the checkpoint with the minimum held-out loss was selected. The gap between training loss (0.59 at the end of training) and held-out loss (0.86) is consistent with a repository-disjoint evaluation; the late upturn in held-out loss indicates that the one-epoch budget approximately exhausts the useful signal for this recipe, and the checkpoint-selection rule guards against the incipient overfitting.

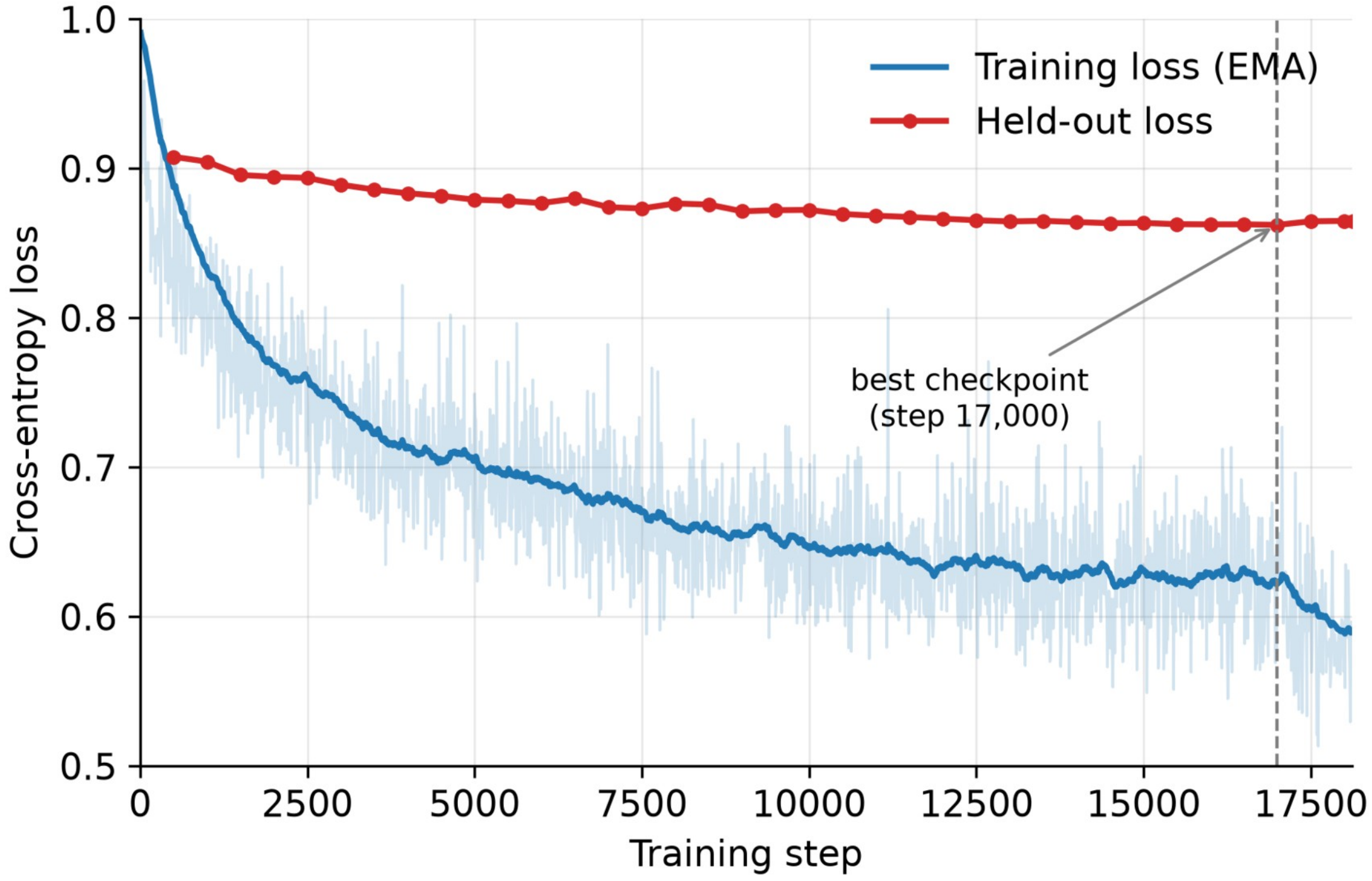

*Figure 2. Training loss and held-out loss over the adaptation run. The dashed line marks the selected checkpoint (step 17,000).*

## 10.4 Evaluation Protocol

The base and adapted models are evaluated under one frozen protocol; the only difference between the two evaluations is the model. Three complementary metric families are used.

*Functional correctness.* Pass@1 — the fraction of tasks whose single greedy completion passes all unit tests — on MultiPL-E (Cassano et al., 2023) for PHP, JavaScript, and TypeScript (161, 161, and 159 tasks respectively), executed with the BigCode evaluation harness, and on HumanEval and its extended-test variant HumanEval+ for Python (164 tasks), scored with the EvalPlus evaluator (Liu et al., 2023). This reproduces the benchmark suite used in the base model's own reporting. Completions are truncated at function boundaries; decoding is greedy with a 512-token limit.

*Held-out language modeling.* Perplexity — the exponentiated mean token-level cross-entropy — per language over the 9,363 held-out files in the four evaluated languages (1.68M evaluated tokens for PHP, 1.59M JavaScript, 0.92M TypeScript, 0.67M Python; the full held-out split comprises 13,827 files across 52 repositories), computed deterministically over fixed-length 1,024-token windows. Because it is computed over millions of tokens without sampling, this is the most statistically stable measurement in the study.

*Held-out code completion.* Two single-line completion tasks on the held-out repositories, with 300 examples per language: FIM infilling (reconstruct a masked line given surrounding context — the trained objective) and next-line completion (continue from the preceding context). Both are scored by exact match and by character-level edit similarity, defined as 1 − Levenshtein distance / max length. The same examples are used for both models, enabling paired comparison.

Statistical treatment: for Python pass@1, where per-task outcomes are recorded for both models, we use McNemar's exact test on discordant pairs; for MultiPL-E, where per-task outcomes were not retained in our runs, so that a paired test could not be applied post hoc, we report Wilson 95% intervals (Table G.3) and an unpaired two-proportion test; perplexity requires no significance test given its determinism and scale. The complete adaptation and evaluation configuration, together with the full per-language result grids and confidence intervals, is given in Appendix G.

## 10.5 Results

| Language | Perplexity | Δ | FIM infill edit similarity | Δ (pts) | Next-line edit similarity | Δ (pts) |
|---|---|---|---|---|---|---|
| PHP | 2.122 → 2.010 | −5.3% | 0.748 → 0.763 | +1.4 | 0.664 → 0.684 | +2.0 |
| JavaScript | 2.003 → 2.052 | +2.4% | 0.819 → 0.803 | −1.5 | 0.731 → 0.726 | −0.5 |
| TypeScript | 2.573 → 2.356 | −8.4% | 0.790 → 0.818 | +2.8 | 0.669 → 0.701 | +3.2 |
| Python | 2.240 → 2.334 | +4.2% | 0.770 → 0.795 | +2.6 | 0.713 → 0.711 | −0.2 |

*Table 5. Held-out measurements before and after adaptation. Δ for perplexity is the relative change (negative is better); Δ for edit similarity is the absolute change in points, that is, the 0–1 metric scaled by 100 (positive is better). Each completion cell aggregates 300 paired examples; perplexity is computed over the token volumes listed in Section 10.4.*

| Benchmark | n | Base | Adapted | Δ | Significance |
|---|---|---|---|---|---|
| MultiPL-E PHP | 161 | 0.497 | 0.404 | −0.093 | p ≈ 0.09 (two-proportion) |
| MultiPL-E JavaScript | 161 | 0.522 | 0.453 | −0.068 | p ≈ 0.22 (two-proportion) |
| MultiPL-E TypeScript | 159 | 0.541 | 0.409 | −0.132 | **p ≈ 0.018** (two-proportion) |
| HumanEval (Python) | 164 | 0.482 | 0.415 | −0.067 | p = 0.071 (McNemar exact) |
| HumanEval+ | 164 | 0.396 | 0.360 | −0.037 | p = 0.345 (McNemar |
| (Python) | | | | | exact) |

*Table 6. Greedy pass@1 before and after adaptation. Δ is the absolute change, computed from unrounded values. Significance: McNemar's exact test on paired per-task outcomes for the Python benchmarks; an unpaired two-proportion test for MultiPL-E, where per-task pairing is unavailable (Section 10.4). Wilson 95% intervals are reported in Table G.3.*

Three findings emerge. First, partially consistent with H1, adaptation improves the modeling and completion of unseen enterprise code in two of the three dominant languages: TypeScript shows the clearest, converging gain — perplexity improves by 8.4% and completion improves on both tasks and both metrics — and PHP shows a more modest gain (perplexity −5.3%, edit similarity up on both tasks), whereas JavaScript — the base model's best-fit language (baseline perplexity 2.00) — shows a small regression on all held-out measures. A plausible post-hoc reading of the ordering is headroom: baseline perplexity is highest on TypeScript (2.57, against 2.00–2.24 for the other languages), exactly where CIDR supplies its third-largest language share. Exact-match results (Table G.1) are more mixed: TypeScript improves on both tasks and Python on FIM infilling, while PHP FIM infilling and both JavaScript tasks decline slightly; edit similarity, which credits near-misses, is the more sensitive of the two metrics for single-line completion. Second, consistent with H2's domain-specificity, Python — the within-training control (included at 5.1% of training tokens but marginal to CIDR's enterprise-web profile) — shows a mixed picture: perplexity degrades by 4.2% and next-line completion is flat, while FIM infilling improves on both metrics, a gain plausibly attributable to transfer of the trained infilling format itself rather than to domain content. Third, functional-correctness scores on docstring-driven benchmarks decline under adaptation, though the decline reaches significance only for TypeScript; the Python declines are not statistically resolvable at $n = 164$ ($p = 0.071$ and $0.345$). We emphasize the scope of these results: absent an equal-budget control adapted on public code of the same languages, they quantify the effect of adaptation on CIDR-derived data, not the differential value of its industrial provenance (Section 10.6).

The pass@1 decline warrants a mechanistic explanation rather than dismissal. The benchmarks measure synthesis of a function body from a natural-language docstring, while CIDR's anonymization confines its redactions to comments and string literals (Section 6) — the very channels carrying the natural-language signal those benchmarks rely on — and the adaptation recipe applies no mitigation against forgetting (no replay of general data). Under one epoch of such training, a shift of probability mass from docstring-conditioned synthesis toward code-conditioned continuation is the expected outcome, and the language with the largest domain gain (TypeScript) correspondingly shows the largest benchmark decline. We report this trade-off transparently: it delineates what a minimal-budget adaptation of a benchmark-competitive base model does and does not provide, and it motivates the replay-based and instruction-preserving recipes that we leave to future work.

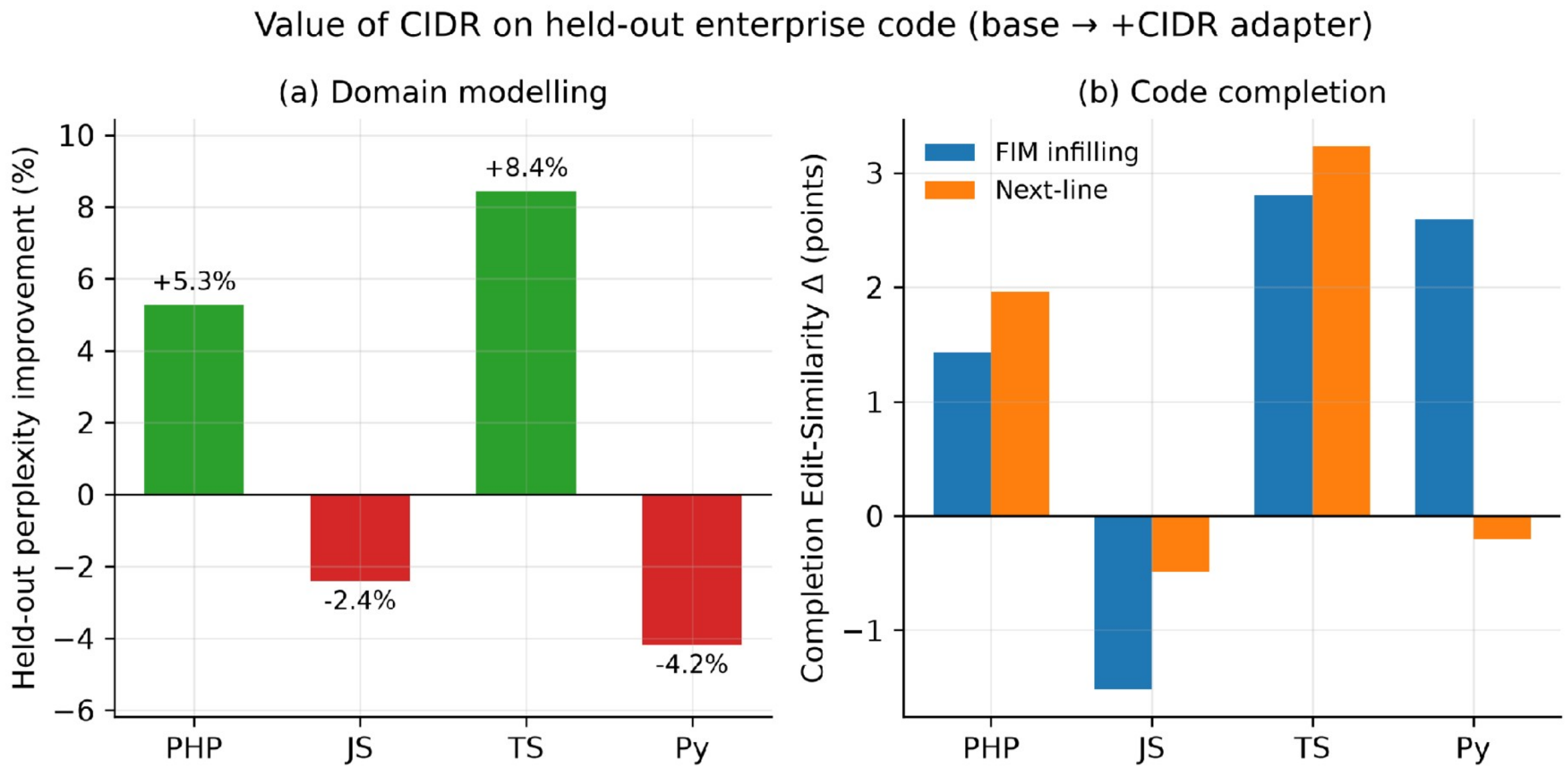

*Figure 3. Effect of adaptation on held-out enterprise code, by language: (a) held-out perplexity improvement; (b) completion edit-similarity change for FIM infilling and next-line completion.*

## 10.6 Threats to Validity

*Attribution.* The study demonstrates that adaptation on CIDR produces a domain-specific gain, but includes no control model trained on an equal token budget of public code; the gains can therefore not be attributed to CIDR's industrial provenance specifically, as opposed to continued pre-training on additional code of matching languages. An equal-budget control is the single most important follow-up experiment.

*Evaluation distribution.* The held-out measurements are computed on repositories held out from CIDR itself. Repository-disjointness excludes memorization, but the evaluation remains in-distribution; performance on enterprise code from unrelated organizations is not measured.

*Statistical power.* With approximately 160 tasks per language, the functional-correctness benchmarks cannot reliably resolve effects of the observed magnitude: only the largest observed difference (TypeScript, −0.132) reaches significance, most pass@1 differences are not individually significant, and single-benchmark point estimates should not be over-read.

*Checkpoint selection.* The checkpoint was selected by minimum loss on a 399-file sample of the same held-out split on which the final perplexities are reported; this can bias the held-out measurements optimistically, although neighboring checkpoints differ by far less than the reported per-language effects.

*Single configuration.* One model, one seed, one recipe, and single-line completion tasks. The completion gains, while consistent across tasks and metrics for TypeScript, are of modest absolute magnitude (2–3 points of edit similarity).

## 11. Quality Assurance

Quality control is applied at each stage of the collection and processing pipeline rather than as a single terminal step, ensuring that errors are identified and addressed as early as possible.

At the metadata stage, automated checks verify that each record is complete, internally consistent, and free of anomalous values — for example, a commit_count of zero in a repository with a non-trivial loc and files value. Records that fail these checks are flagged for manual inspection before the repository advances to the code review stage.

At the anonymization stage, the pipeline enforces a hard gate check as described in Section 6.2: the output bundle is automatically re-scanned after redaction, and any remaining findings cause the repository to be withheld pending manual resolution. In addition, a random sample of 60% of accepted repositories undergoes a further manual audit after anonymization, in which a member of the curation team inspects the redacted output directly to verify that no sensitive content remains and that the functional structure of the codebase has been preserved.

At the partner agreement stage, each contributing organization signed a data sharing agreement prior to submission. Beyond the anonymization and access-control provisions described in Section 8, these agreements include explicit contractual representations by the partner that the contributed repositories constitute genuine, human-authored software developed within their organization, and that the contribution does not violate applicable privacy or intellectual property obligations. This contractual layer complements the automated rejection criteria described in Section 5.3 by establishing legal accountability for the authenticity of the submitted material.

Annotator guidelines used during manual code review are not published. Disclosure would create an incentive for contributing partners to artificially adapt submitted repositories to satisfy the stated criteria — for example, by selectively removing low-quality files, restructuring code to superficially resemble the expected architectural patterns, or staging commit histories to mask the true development process. The value of the manual review stage depends precisely on its capacity to assess repositories as they exist in practice, without anticipatory modification; publishing the evaluation rubric would undermine this by shifting partner behaviour from genuine contribution to criterion-targeted presentation. The most frequent categories of rejection encountered during the review process are documented in Section 5.3 and serve as the primary empirical record of systematic quality issues observed across submissions.

## 12. Intended Use Cases and Limitations

### 12.1 Intended Use Cases

- Pre-training and fine-tuning of code language models
- Software defect prediction and code quality research
- Developer behaviour and software evolution studies
- Evaluation of static analysis and code review tools
- Construction of SWE-bench-style evaluation benchmarks for autonomous coding agents, leveraging the full version control history to derive issue–patch pairs and repository-grounded tasks against which agent capabilities can be assessed in realistic industrial settings
- Reinforcement learning from verifiable rewards (RLVR), where the repository structure and associated test suites provide a ground-truth execution environment for generating and verifying reward signals during agent training

### 12.2 Known Limitations

- Language coverage reflects the domains of contributing partners and may not generalize to all programming ecosystems.
- Anonymization may affect the utility of commit messages and documentation strings for natural language tasks.
- The dataset reflects the practices and conventions of the contributing organizations and should not be treated as a universal sample of software industry practices.
- The engineering-practice attributes of Section 9.5 are assigned by rule-based static detection (Section 4.3.2); they measure the presence of tooling artifacts, not the operational practice of the contributing team, and may misclassify unconventional setups.

## 13. Conclusion

CIDR is a large-scale, industry-sourced software repository dataset constructed through a structured collection, curation, and anonymization pipeline developed specifically for this purpose. By partnering directly with industrial organizations, we have been able to assemble a dataset that complements existing open-source collections and provides access to software artifacts that would otherwise be inaccessible to the research community. Unlike prior work that relies on publicly available repositories, CIDR reflects the kinds of production codebases that professional software development teams actively maintain — making it particularly well-suited for research on code intelligence, developer tooling, and software quality in realistic enterprise contexts.

The dataset is accompanied by a transparent description of every stage of the collection and processing pipeline, including the metadata extraction methodology, the two-stage selection and quality criteria, the deterministic anonymization procedure, and the data management infrastructure developed to support the collection effort at scale. We hope that this transparency will facilitate critical assessment of the dataset's properties and support reproducible research built upon it.

Looking ahead, we plan to extend CIDR along three directions. First, we will continue to expand the contributor base by onboarding additional industrial partners, with particular attention to domains and technology stacks that are currently underrepresented in the dataset. Second, we aim to broaden programming language coverage in response to the linguistic profile of incoming submissions, prioritising languages that are prevalent in industrial software but absent from existing public collections. Third, and most substantively, we are developing an automated platform for end-to-end repository ingestion, processing, and delivery. This platform will systematise the submission, anonymization, and quality assurance steps that are currently partially manual, reduce the operational overhead of onboarding new partners, and enable continuous dataset updates as additional repositories are contributed over time. We anticipate that this infrastructure will meaningfully lower the barrier to industrial data collection and serve as a reusable foundation for future dataset efforts in the software engineering research community.

# Appendix

## A. Extended Dataset Statistics

### Table A.1 — Full Language Distribution

*Language distribution across the 4,225 cataloged repositories, sorted by LoC-weighted share. Estimated LoC per language is computed as the per-repository language share multiplied by the repository's raw LoC, summed over repositories. The taxonomy excludes the markup, data, and documentation formats enumerated by the earlier* cloc*-based count, and languages below a 1% within-repository share are omitted (Section 4.3.2).*

| **Language** | **LoC share (%)** | **Estimated LoC (raw)** | **Repositories** |
|---|---|---|---|
| *PHP* | *40.52* | *337,323,346* | *1004* |
| *JavaScript* | *27.93* | *232,478,641* | *1690* |
| *TypeScript* | *5.85* | *48,727,375* | *839* |
| *C++ Header* | *4.08* | *33,981,204* | *30* |
| *C#* | *3.71* | *30,882,128* | *161* |
| *Python* | *3.51* | *29,229,116* | *541* |
| *Ruby* | *1.76* | *14,645,223* | *59* |
| *SQL* | *1.44* | *12,019,313* | *204* |
| *C Header* | *1.35* | *11,264,850* | *168* |
| *Java* | *1.21* | *10,112,916* | *361* |
| *Dart* | *1.16* | *9,614,653* | *151* |
| *Swift* | *1.15* | *9,538,805* | *310* |
| *Go* | *0.97* | *8,046,886* | *66* |
| *Kotlin* | *0.96* | *7,971,265* | *259* |
| *C* | *0.88* | *7,296,161* | *19* |
| *Smarty Template* | *0.55* | *4,594,174* | *93* |
| *C++* | *0.51* | *4,277,062* | *61* |
| *Vue* | *0.48* | *4,025,110* | *154* |
| *Objective C* | *0.4* | *3,318,489* | *148* |
| *Ruby HTML* | *0.23* | *1,896,603* | *16* |
| *Twig Template* | *0.2* | *1,630,709* | *91* |
| *Razor* | *0.15* | *1,207,849* | *25* |
| *CoffeeScript* | *0.13* | *1,100,292* | *13* |
| *Shell* | *0.12* | *1,002,281* | *543* |
| *Blade template* | *0.09* | *721,586* | *78* |
| *Assembly* | *0.08* | *688,664* | *1* |
| *JSX* | *0.08* | *642,707* | *60* |
| *Makefile* | *0.06* | *524,945* | *323* |
| *Visual Basic* | *0.05* | *395,067* | *1* |
| *Alex* | *0.04* | *344,875* | *1* |
| *TypeScript Typings* | *0.04* | *344,808* | *65* |
| *Gradle* | *0.04* | *332,118* | *281* |
| *Perl* | *0.04* | *306,018* | *5* |
| *Objective C++* | *0.03* | *282,738* | *24* |
| *Groovy* | *0.02* | *201,672* | *14* |
| *Cython* | *0.02* | *191,067* | *2* |
| *JavaServer Pages* | *0.02* | *189,558* | *2* |
| *Terraform* | *0.02* | *154,826* | *23* |
| *Solidity* | *0.01* | *118,913* | *12* |
| *Batch* | *0.01* | *107,802* | *195* |
| *QML* | *0.01* | *85,530* | *5* |
| *Basic* | *0.01* | *83,242* | *1* |
| *Autoconf* | *0.01* | *67,998* | *2* |
| *BASH* | *0.01* | *57,740* | *124* |
| Dockerfile | *0.01* | *53,033* | *428* |
| *m4* | *0.01* | *48,691* | *1* |
| *CMake* | *0.01* | *47,671* | *19* |
| *Jinja* | *0.01* | *45,166* | *13* |
| *Cuda* | *0.0* | *33,017* | *1* |
| *Astro* | *0.0* | *24,652* | *3* |
| *Lua* | *0.0* | *23,483* | *9* |
| *ASP.NET* | *0.0* | *20,083* | *2* |
| *TemplateToolkit* | *0.0* | *19,308* | *1* |
| *Go Template* | *0.0* | *19,200* | *6* |
| *Freemarker Template* | *0.0* | *16,736* | *10* |
| *Prolog* | *0.0* | *12,577* | *71* |
| *Rust* | *0.0* | *7,045* | *3* |
| *Snakemake* | *0.0* | *5,474* | *4* |
| *Svelte* | *0.0* | *4,218* | *3* |
| *Bru* | *0.0* | *4,011* | *1* |
| *Powershell* | *0.0* | *3,208* | *4* |
| *HCL* | *0.0* | *2,387* | *14* |
| *Tact* | *0.0* | *2,001* | *2* |
| *Gemfile* | *0.0* | *1,389* | *27* |
| *IDL* | *0.0* | *1,197* | *1* |
| *GLSL* | *0.0* | *1,187* | *1* |
| *Rakefile* | *0.0* | *854* | *5* |
| *Treetop* | *0.0* | *832* | *2* |
| *Flow9* | *0.0* | *746* | *1* |
| *Raku* | *0.0* | *553* | *2* |
| *Mako* | *0.0* | *130* | *4* |
| *Jenkins Buildfile* | *0.0* | *119* | *1* |
| *Bazel* | *0.0* | *72* | *3* |
| *AWK* | *0.0* | *55* | *1* |
| *Android Interface Definition Language* | *0.0* | *20* | *1* |

## Table A.2 — Full Descriptive Statistics for Size Metrics

*Descriptive statistics for the principal size and quality metrics across the 4,225 cataloged repositories. Percentiles P10–P95 are reported alongside mean and standard deviation; metric definitions are given in Section 4.3.2.*

| Metric | sum | mean | std | min | P10 | P25 | P50 | P75 | P90 | P95 | max |
|---|---|---|---|---|---|---|---|---|---|---|---|
| Total LOC (raw) | 832,425,431 | 197,024 | 890,852 | 3 | 680.8 | 2,541 | 14,157 | 67,088 | 337,270 | 884,217 | 22,960,732 |
| Logical LOC | 580,658,442 | 137,434 | 726,464 | 2 | 521 | 1,856 | 10,384 | 48,484 | 230,711 | 565,704 | 19,653,113 |
| Auto-Generated LOC | 27,095,927 | 6,413 | 147,438 | 0 | 0 | 0 | 0 | 213 | 3,257 | 9,836 | 8,914,800 |
| Source File Count | 4,528,212 | 1,072 | 4,099 | 1 | 14 | 43 | 160 | 597 | 2,102 | 4,302 | 95,167 |
| Commit Volume | 1,599,144 | 378.5 | 1,368 | 0 | 1.4 | 8 | 46 | 232 | 809 | 1,505 | 26,333 |
| Total PRs | 316,607 | 74.94 | 346.95 | 0 | 0 | 0 | 1 | 27 | 152.6 | 320.6 | 9,765 |
| Duplication Rate | 298.79 | 0.07 | 0.13 | 0 | 0 | 0 | 0.0122 | 0.0775 | 0.229 | 0.3675 | 0.86 |
| Documentation Density | 534.3 | 0.13 | 0.21 | 0 | 0 | 0.0024 | 0.0357 | 0.1479 | 0.3633 | 0.6354 | 1 |
| Avg. Fn. Length (LOC) | 53,623 | 12.69 | 9.07 | 0 | 2.994 | 7.7 | 11.48 | 16.17 | 21.942 | 27 | 113 |

## Figure A.1 — Language Distribution by LoC Share and Repository Count

*Top 15 programming languages in CIDR by (left) share of total lines of code and (right) number of repositories in which the language occurs, across the 4,225 cataloged repositories.*

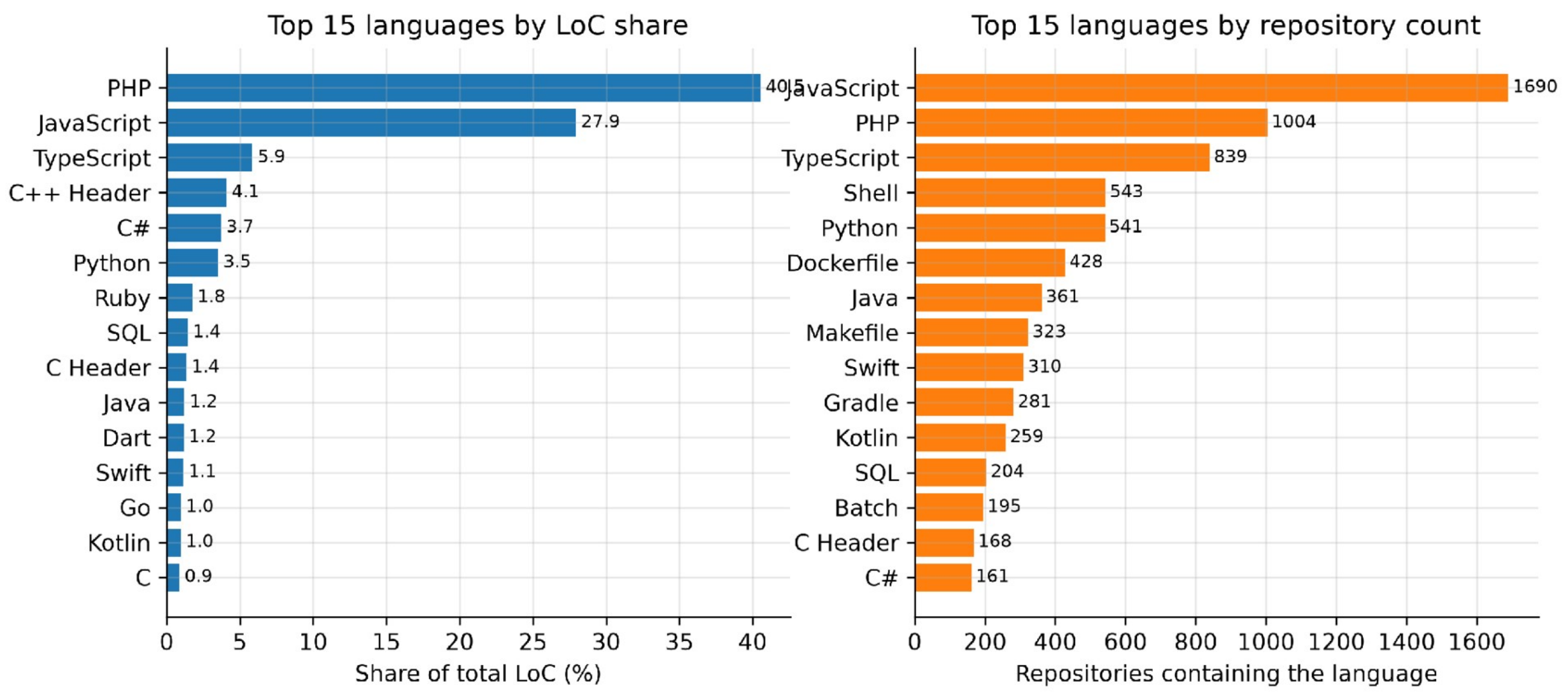

## Figure A.2 — Primary Language Distribution by Repository Count

*Distribution of cataloged repositories by primary programming language, defined as the language with the highest share of code lines within each repository. Languages accounting for less than 2% of repositories are grouped into "Other".*

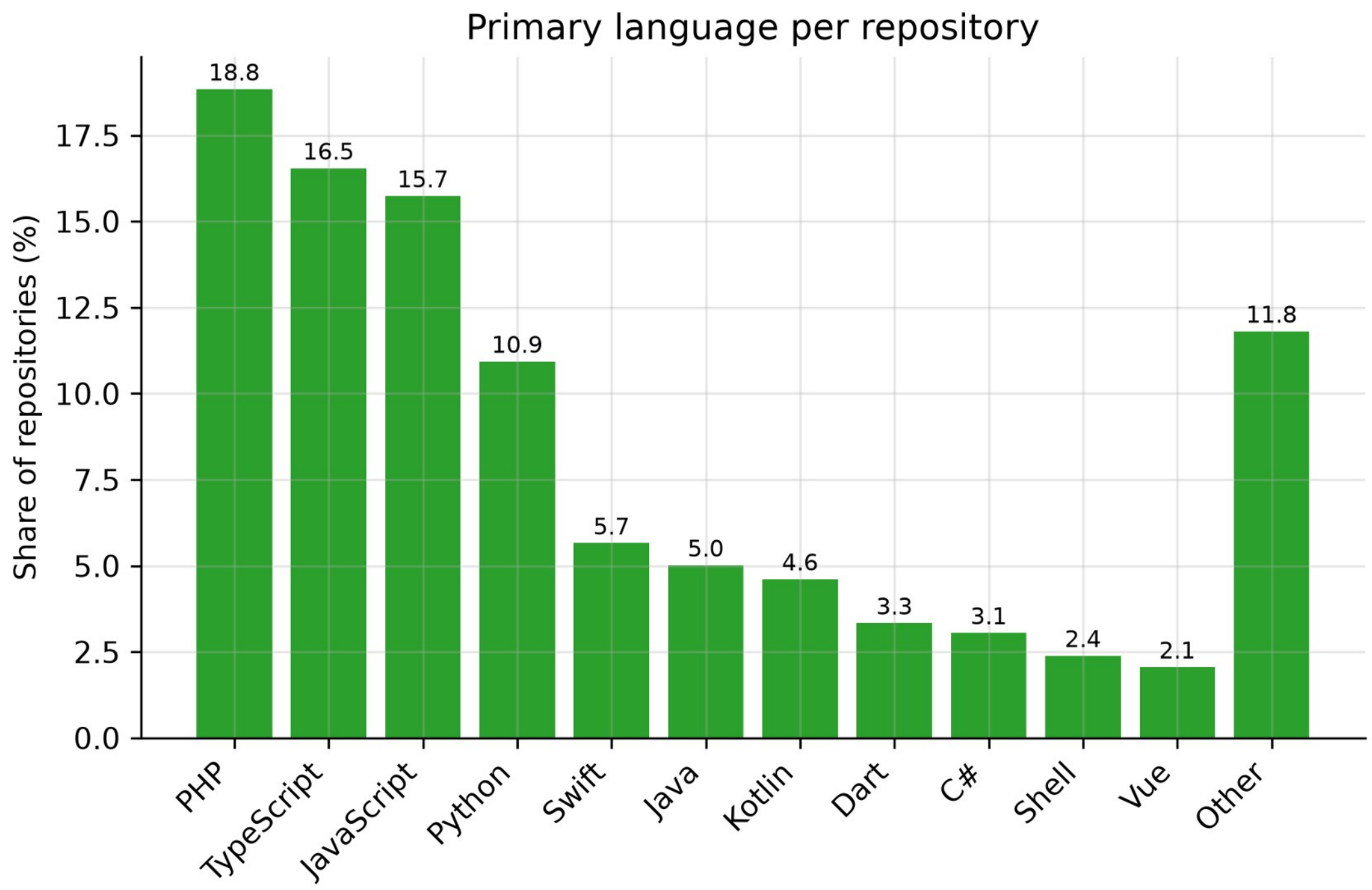

### Figure A.3 — Repository Distribution by Raw LoC Bucket

*Number of cataloged repositories in each raw-LoC size bracket. Bucket boundaries: <1K, 1K–5K, 5K–20K, 20K–100K, 100K–500K, >500K raw lines of code.*

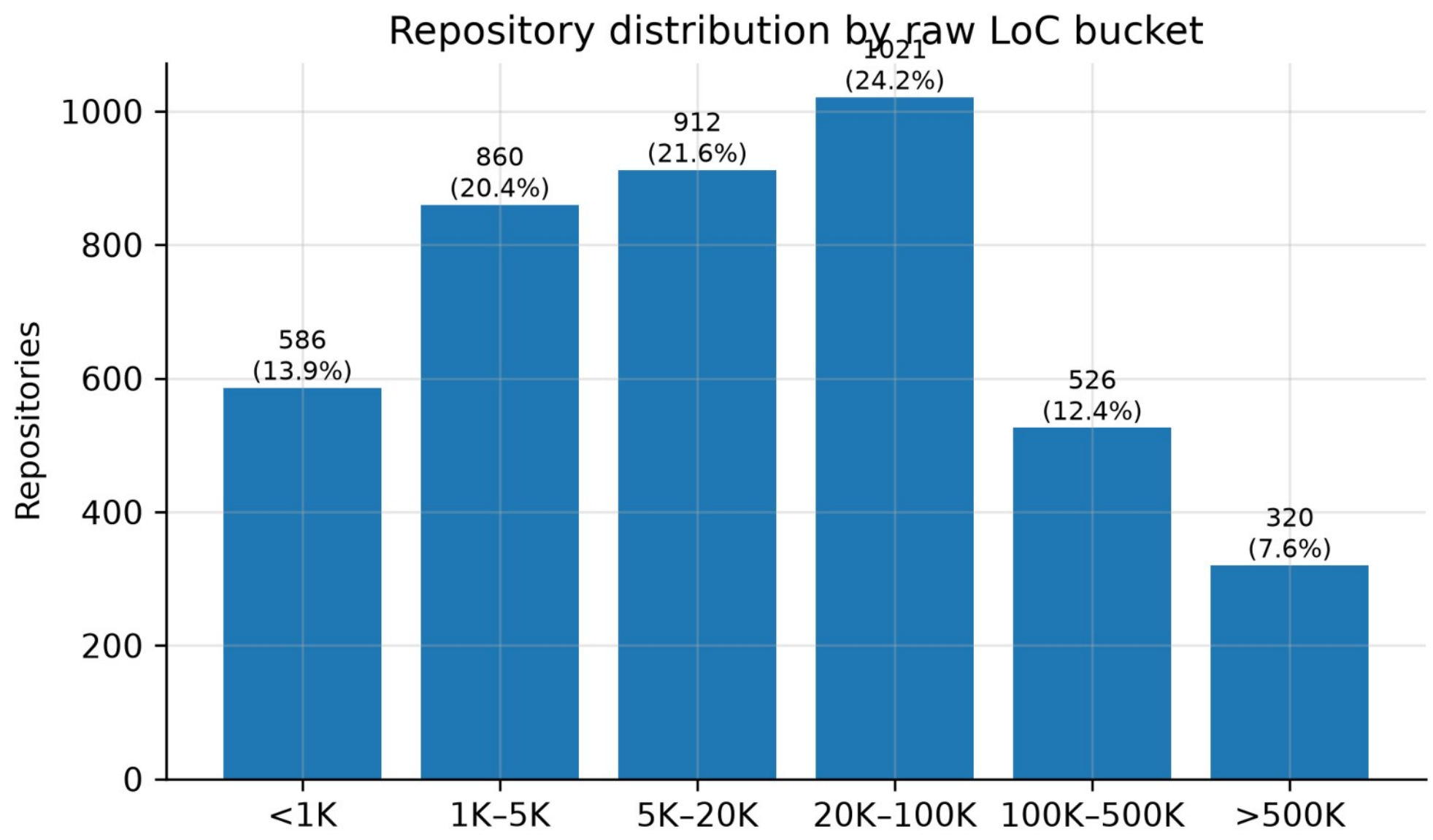

### Figure A.4 — Code Quality Metric Distributions

*Distributions of three code quality metrics across cataloged repositories. Left: duplication rate (*jscpd*); centre: documentation density (share of documented functions); right: average function length in lines (Tree-sitter). Red dashed lines indicate medians (Section 4.3.2).*

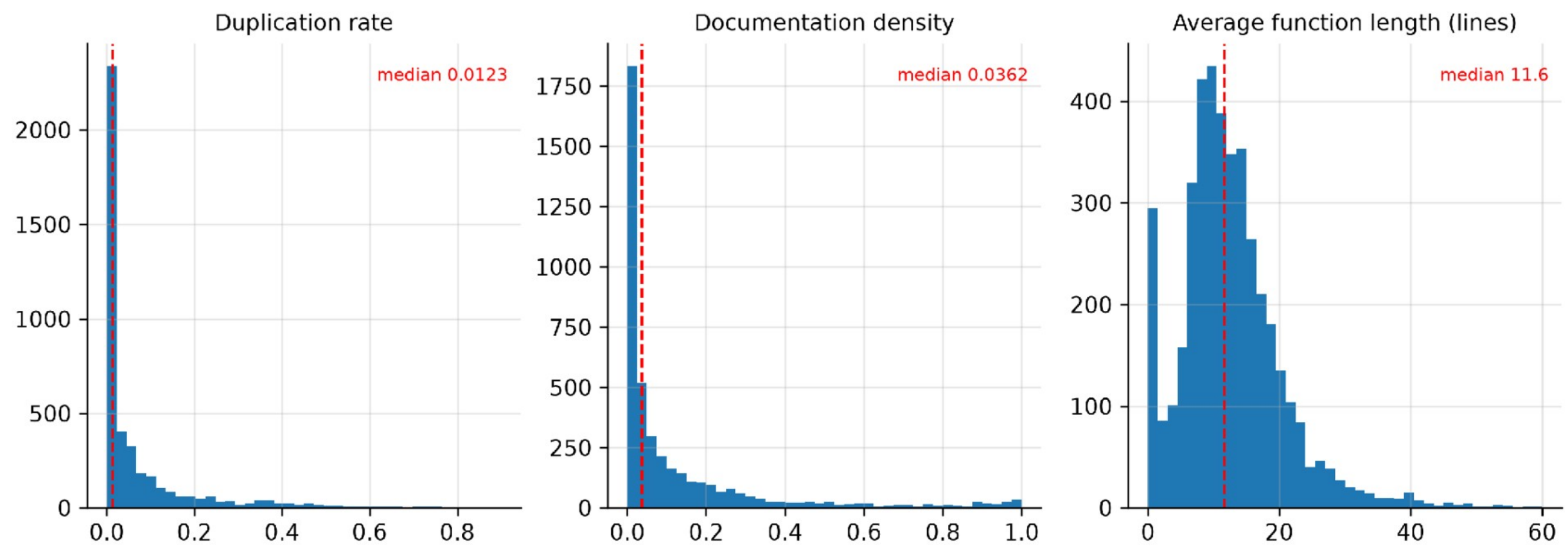

## Figure A.5 — Distribution of Repository First-Commit Year

*Number of accepted repositories by year of first commit, covering the period 2011–2026. Repositories with missing or unparseable timestamps are excluded. Initial collection campaign (2,545 accepted of 4,449 submitted).*

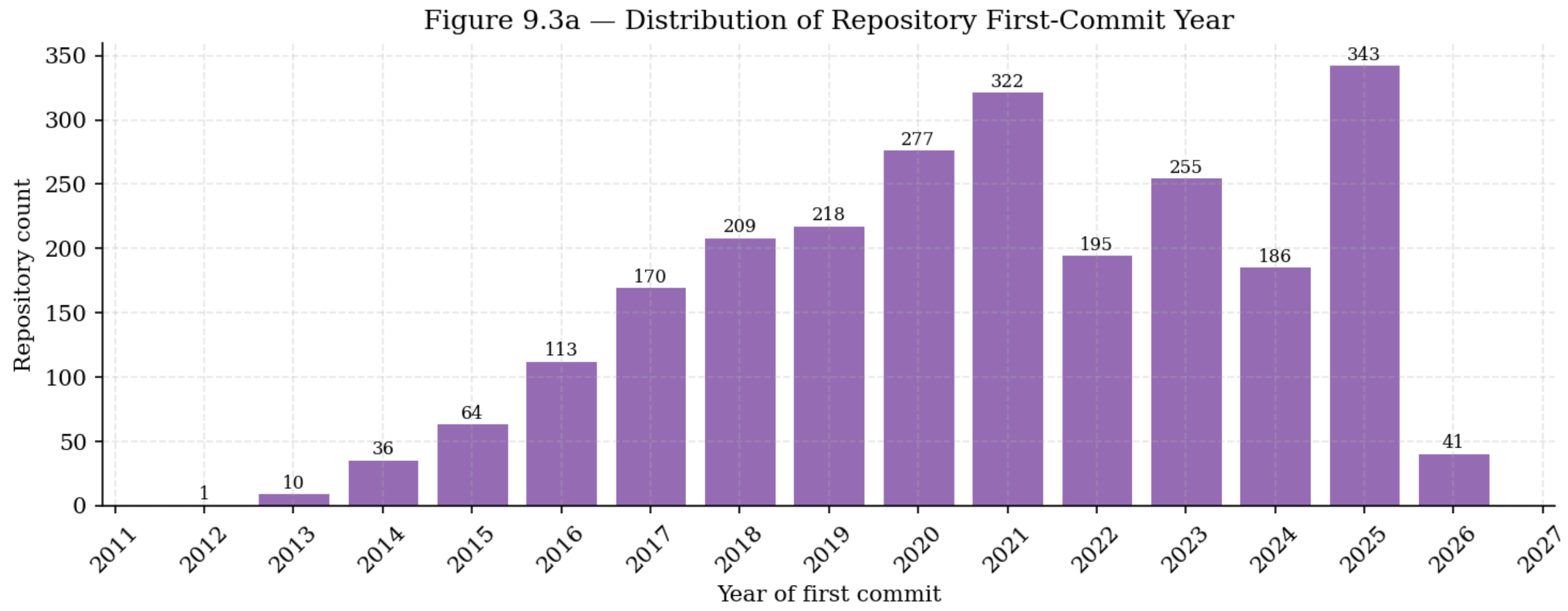

## Figure A.6 — Repository Age vs. Commit Count

*Scatter plot of accepted repositories by year of first commit (x-axis) and $log_{10}$(commit count) (y-axis). Point colour encodes $log_{10}$(loc). Initial collection campaign (2,545 accepted of 4,449 submitted).*

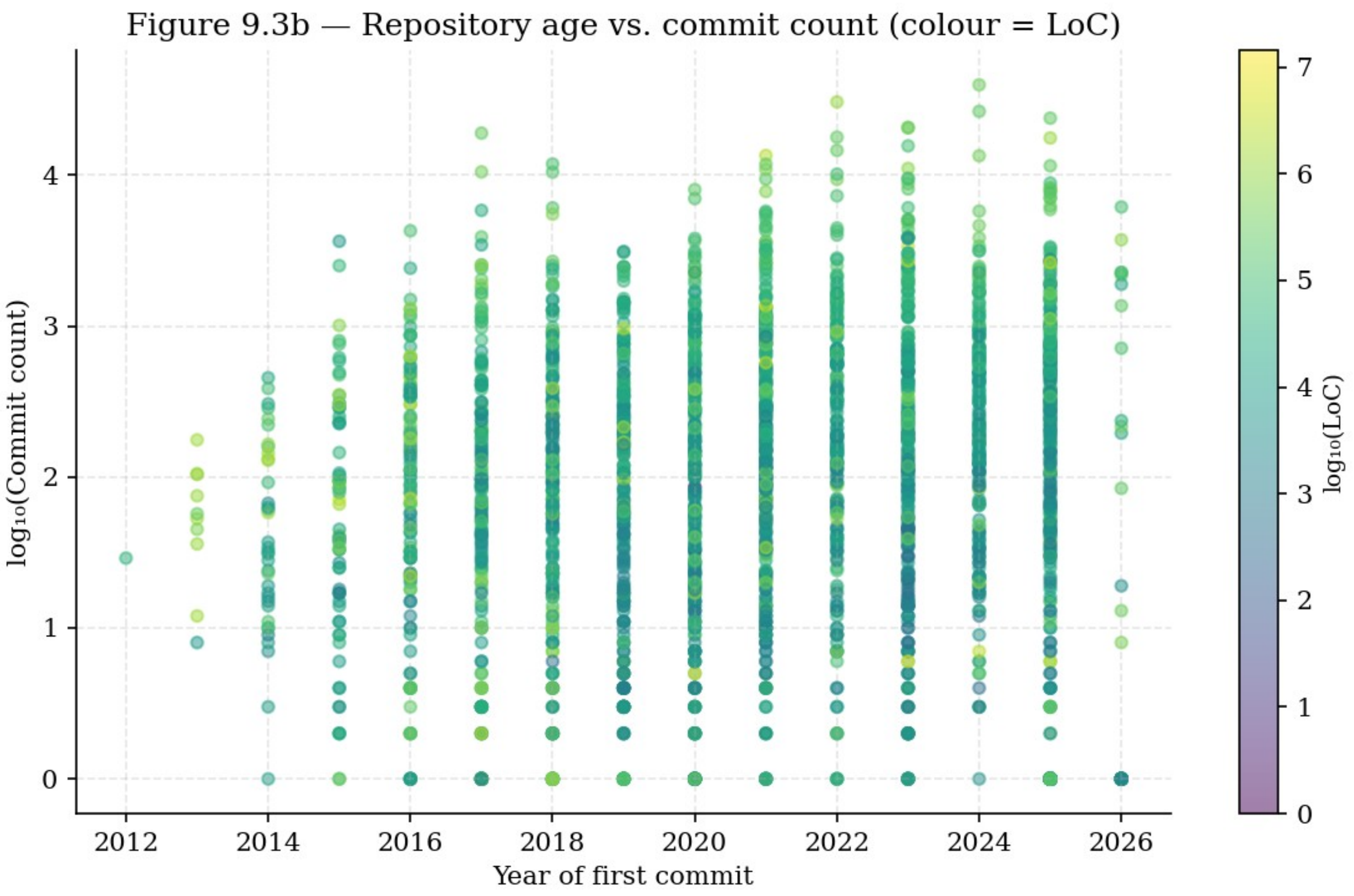

### Figure A.7 — Repository Status Distribution

*Distribution of all 4,449 submitted repositories by lifecycle status at the time of dataset compilation. Percentages are relative to total submissions. Initial collection campaign (2,545 accepted of 4,449 submitted).*

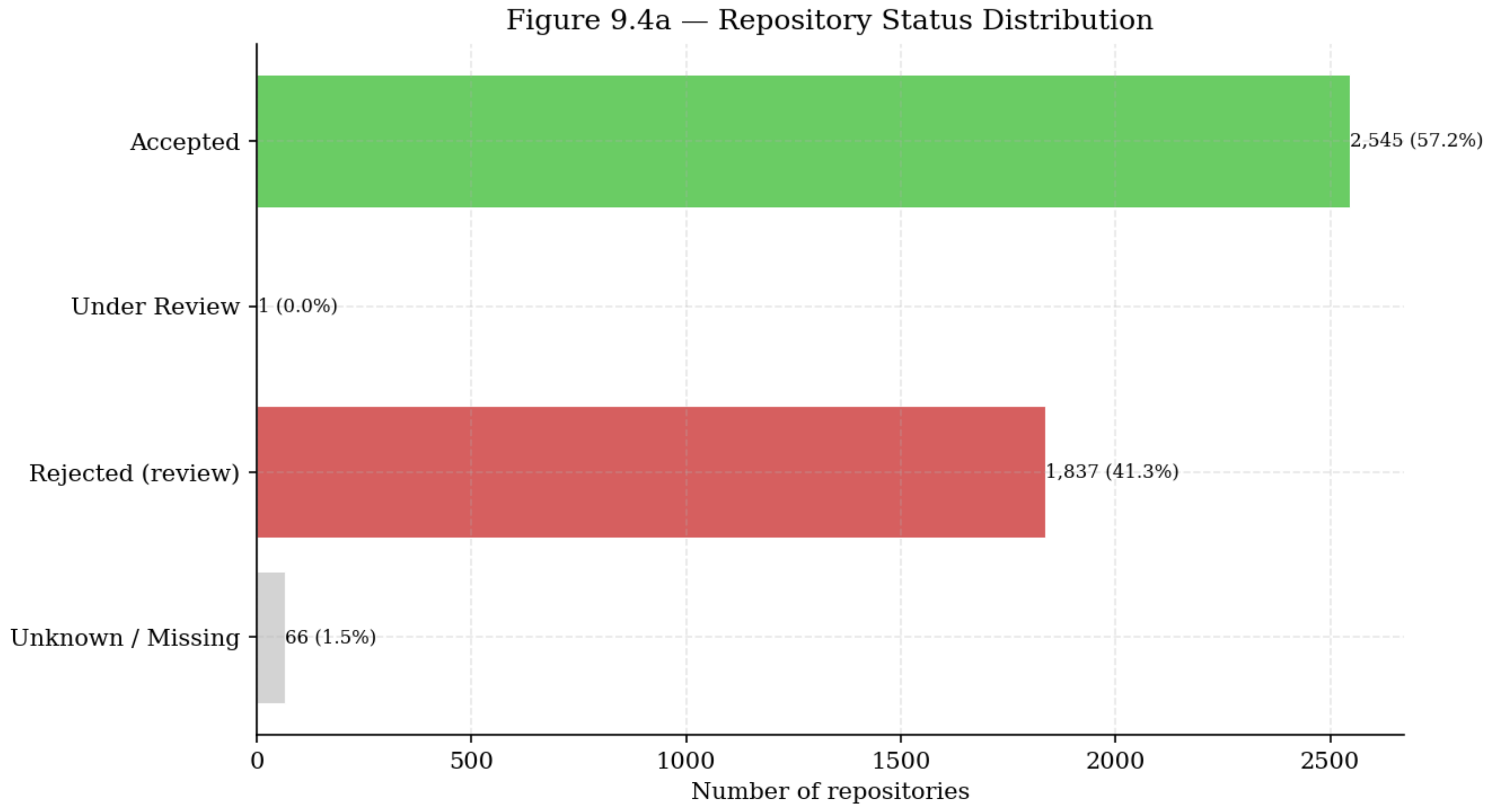

## Figure A.8 — Distribution of Rejection Reasons

*Number of repositories rejected at the code review stage per rejection category (Section 5.3). Categories are not mutually exclusive. Initial collection campaign (2,545 accepted of 4,449 submitted).*

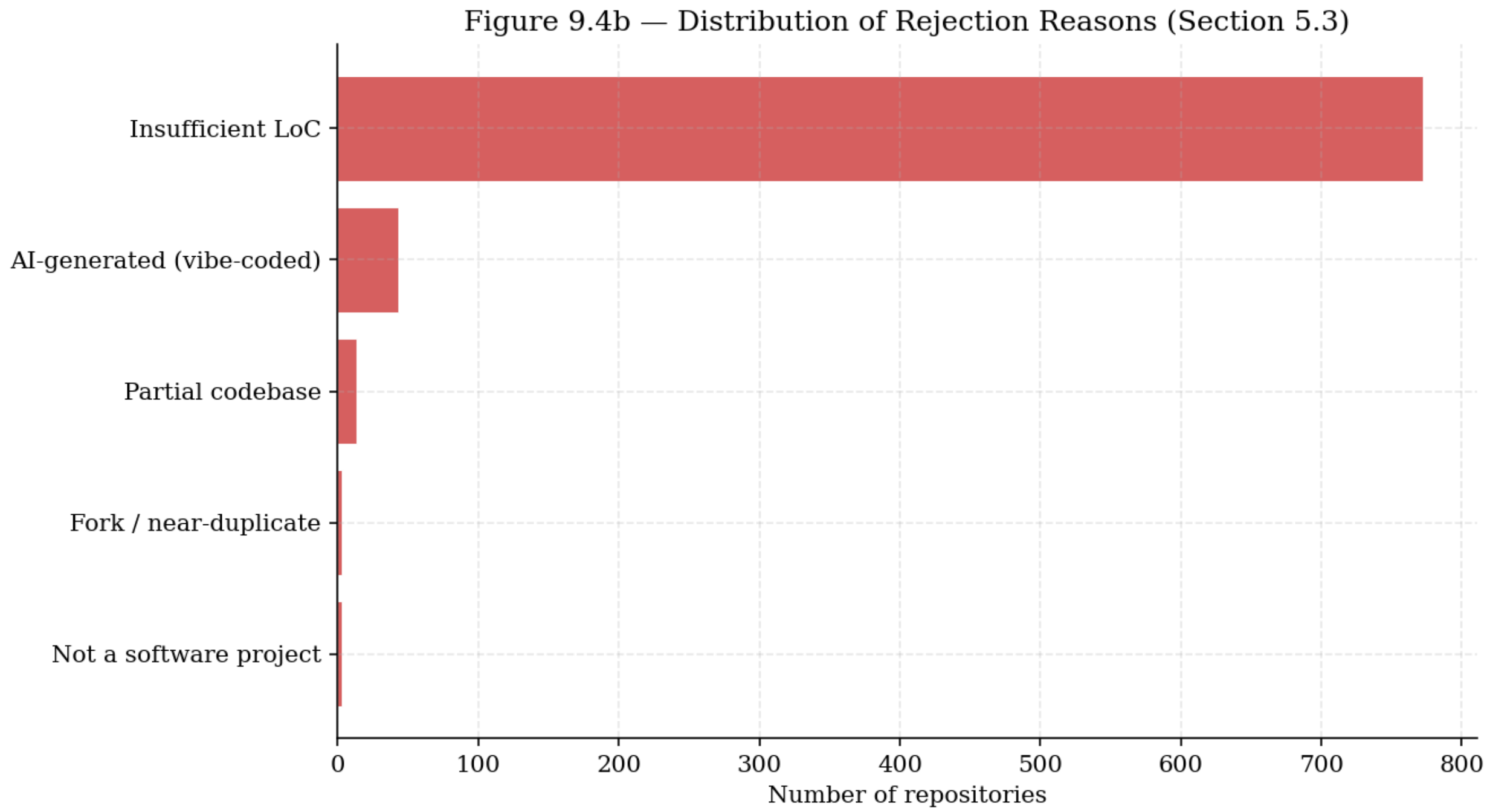

## Figure A.9 — Engineering Practice Attribute Distributions

*Distributions of the seven engineering-practice attributes across the 4,225 cataloged repositories (Section 9.5); attribute definitions and detection rules are given in Section 4.3.2 and Appendix H.*

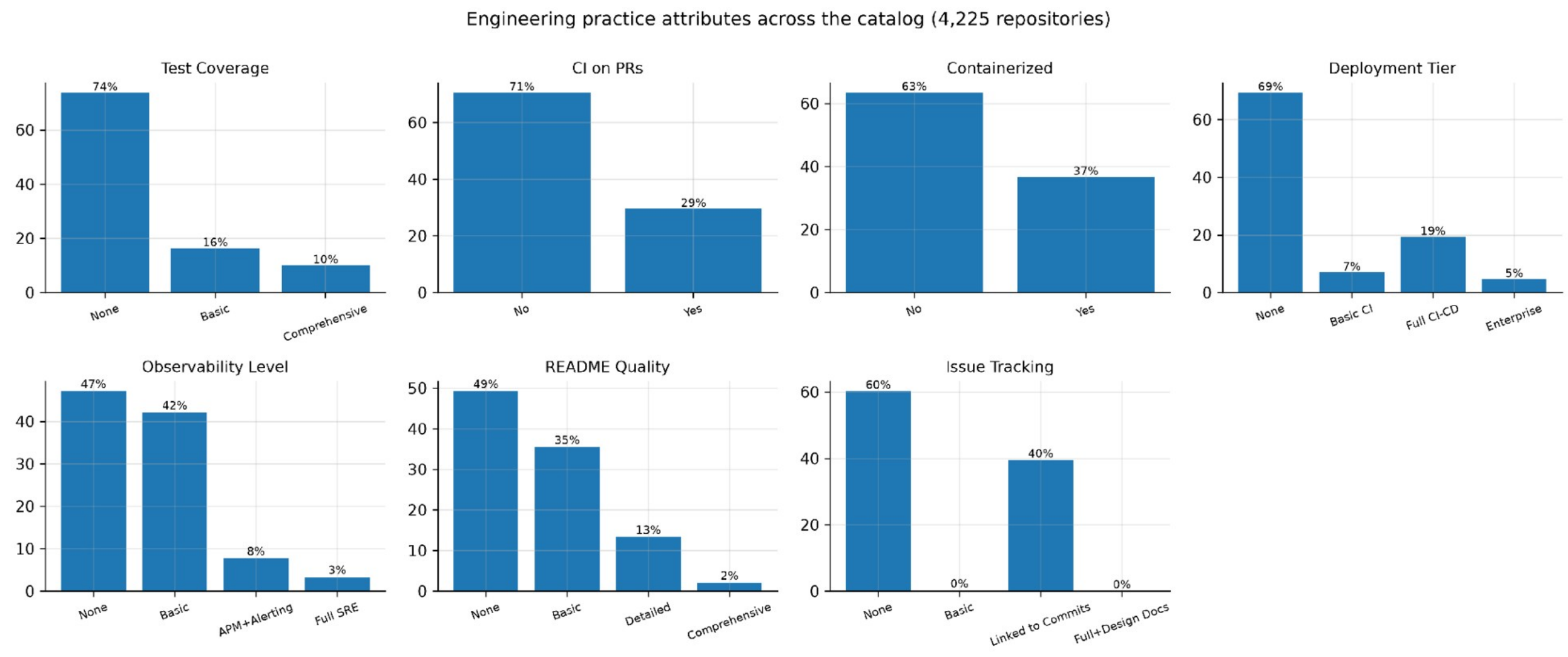

## B. Anonymization Rules Reference

*Full enumeration of detection patterns applied by the anonymization utility across the five detector classes described in Section 6.2. For each detector, the table lists the pattern name or entity type, the detection mechanism, the replacement mask applied, and the pipeline stages in which the detector is active.*

*Stage abbreviations: WT = working tree scan (steps 3/5); HM = commit metadata scan (steps 6/8); HB = history blob scan (steps 6b/8b); HR = history rewrite callbacks (step 7).*

| Detector | Pattern / Entity type | Detection mechanism | Replacement mask | Active stages |
|---|---|---|---|---|
| *SecretsDetector* | API keys, tokens, passwords, private keys, AWS credentials, GitHub PATs, JWT tokens (full gitleaks rule set) | gitleaks detect --no-git; rule-based pattern matching against 100+ known secret formats | REDACTED_{hash12} | WT, HM |
| *RegexPIIDetector* | Email address | Regex: [a-zA-Z0-9_.+-]+@[a-zA-Z0-9-]+\.[a-zA-Z0-9-.]+ | user_{hash12}@example.com | WT, HM, HB, HR |
| *RegexPIIDetector* | Phone number (E.164) | Regex: E.164 international phone format | +0000000000 | WT, HM, HB, |
| | | | | HR |
| *RegexPIIDetector* | IPv4 address | Regex: dotted-decimal IPv4 | 192.0.2.{1–254} | WT, HM, HB, HR |
| *RegexPIIDetector* | JWT token | Regex: [A-Za-z0-9-_]+ \.[A-Za-z0-9-_]+\.[A-Za-z0-9-_]+ | REDACTED_{hash12} | WT, HM, HB, HR |
| *RegexPIIDetector* | HTTPS URL | Regex: https?://[^\s]+ | [url:{hash12}] | WT, HM, HB, HR |
| *RegexPIIDetector* | Custom patterns (regex/pii_patterns.yaml) | User-defined Python regex; compiled with re.compile(pattern); no flags by default; (?i) for case-insensitive | [name:{hash12}] (pattern name used as mask label) | WT, HM, HB, HR |
| *DictionaryDetector* | Partner codenames (dict/codenames.txt) | Aho-Corasick multi-pattern search over lowercased content; O(n) in text length; one match per term regardless of frequency | [codename:{hash12}] | WT, HM, HB |
| *DictionaryDetector* | Client names (dict/clients.txt) | Aho-Corasick multi-pattern search; case-insensitive; terms may contain spaces and special characters | [client:{hash12}] | WT, HM, HB |
| *DictionaryDetector* | Organisational unit names (dict/orgs.txt) | Aho-Corasick multi-pattern search; case-insensitive | [org:{hash12}] | WT, HM, HB |
| *DictionaryDetector* | Internal domains (dict/domains.txt) | Aho-Corasick string matching (detects domain name string anywhere in content) | {hash8}.example.invalid | WT, HM, HB |
| *DictionaryDetector* | Custom terms (any | Aho-Corasick multi- | [term:{hash12}] | WT, |
| | dict/*.txt) | pattern search; file stem used as dictionary name in findings | | HM, HB |
| *EndpointDetector* | Private IP — Class A (RFC 1918) | Rule-based: 10.0.0.0/8 address range | 192.0.2.{1–254} | WT, HM, HB |
| *EndpointDetector* | Private IP — Class B (RFC 1918) | Rule-based: 172.16.0.0/12 address range | 192.0.2.{1–254} | WT, HM, HB |
| *EndpointDetector* | Private IP — Class C (RFC 1918) | Rule-based: 192.168.0.0/16 address range | 192.0.2.{1–254} | WT, HM, HB |
| *EndpointDetector* | Internal TLD — .internal | Structural suffix matching on domain names | {hash8}.example.invalid | WT, HM, HB |
| *EndpointDetector* | Internal TLD — .corp | Structural suffix matching on domain names | {hash8}.example.invalid | WT, HM, HB |
| *EndpointDetector* | Internal TLD — .local | Structural suffix matching on domain names | {hash8}.example.invalid | WT, HM, HB |
| *EndpointDetector* | Internal TLD — .lan | Structural suffix matching on domain names | {hash8}.example.invalid | WT, HM, HB |
| *EndpointDetector* | Internal TLD — .intra | Structural suffix matching on domain names | {hash8}.example.invalid | WT, HM, HB |
| *EndpointDetector* | Partner-specific domains (dict/domains.txt) | Structural suffix matching using custom domain list (dual use with DictionaryDetector) | {hash8}.example.invalid | WT, HM, HB |
| *NERDetector* | Person name (PER) | Multilingual BERT transformer (Babelscape/wikineural-multilingual-ner) or GLiNER zero-shot; | Person_{hash12} | WT, HM |
| | | min. confidence score configurable (default 0.5); entities shorter than 3 characters discarded; long texts split into overlapping chunks | | |
| *NERDetector* | Organisation name (ORG) | Same transformer model as PER; ORG label mapped to ORG_NAME category; same minimum score and length thresholds | Org_{hash12} | WT, HM |
| *History rewrite* | Commit author name | git-filter-repo name_callback; applied to every commit on every branch and tag | Author_{hash12} | HR |
| *History rewrite* | Commit author email | git-filter-repo email_callback; applied to every commit on every branch and tag | author_{hash12}@example.invalid | HR |
| *History rewrite* | Commit message body | git-filter-repo message_callback; RegexPII patterns applied as byte-level regex to message text | [name:{hash12}] | HR |
| *History rewrite* | Blob content — PII patterns | git-filter-repo blob_callback; all patterns from regex/pii_patterns.yaml applied to text blob content | [name:{hash12}] | HR |
| *History rewrite* | Denied files | git-filter-repo filename_callback; files matching any deny_glob removed from every commit in the rewritten history | File deleted from history | HR |

*Notes. (1)* {hash12} = HMAC-SHA256(salt, original_value).hexdigest()[:12]*;* {hash8} *= first 8 characters. The salt is operator-controlled and never included in the output bundle. (2)* SecretsDetector *and* NERDetector *are excluded from history blob scanning (steps 6b/8b): gitleaks subprocess invocation and ML inference per blob would be computationally prohibitive across large object stores. History blob coverage relies exclusively on* RegexPIIDetector*,* DictionaryDetector*, and* EndpointDetector*. (3) For code files in the working tree, all detectors operate only within tree-sitter zones — comments, string literals, and docstrings — leaving identifiers and code structure unmodified. For configuration and documentation files, entire file content is scanned. (4) Zone-based restriction is not applied to commit metadata or historical blob scanning; both surfaces are scanned in full.*

## C. Partner Portal — Feature Summary

*Description of the functionality provided through the partner self-service portal (Section 7.3).*

The partner self-service portal is a web-based interface through which contributing organizations manage their repository submissions and interact with the curation team. Access is granted by the curation team at onboarding; partners authenticate using a hash token rather than a conventional username, which avoids exposing internal account identifiers in external communications. The portal is accessible at /partner/rows/ following login and is intentionally scoped to the partner's own data: records belonging to other partners are neither visible nor reachable, and attempts to access them by UUID return a 404 response without disclosing the existence of the record.

### Table C.1 — Partner Portal Functionality Summary

| Feature | Description | Notes |
|---|---|---|
| *Repository list* | Interactive table displaying all repository records belonging to the authenticated partner. Supports column sorting, per-column filtering, column reordering, and a filtered/total record counter. | Columns with partner_visible = False are completely absent from the interface. |
| *Metadata entry form* | Guided form for submitting a new repository record. Each field is rendered according to its declared type: dropdown selectors for enumerated attributes, date pickers for temporal fields, validated URL inputs for repository addresses, checkboxes for boolean fields, and | Required fields are enforced. Duplicate records trigger a warning before saving. Each submission is assigned a UUID (repo_id) automatically. |
| | text areas for free-form and JSON fields. | |
| *Record editing* | Full edit form for an existing record, pre-populated with current values. Fields with partner_editable = False are displayed but locked. | Only records belonging to the authenticated partner are editable. |
| *Inline cell editing* | Direct in-table editing without navigating to the edit form. A double-click on any editable cell opens an input; the value is validated server-side and saved via AJAX without page reload. Validation errors are surfaced inline. | Available only for columns with partner_editable = True. |
| *Lifecycle status tracking* | Each repository record displays its current lifecycle status (submitted, under review, accepted, anonymized, packaged, delivered) and the full history of status transitions, allowing partners to monitor the progress of their submissions. | Status transitions are managed exclusively by the curation team. |
| *CSV import* | Bulk upload of repository records via CSV file. Supports creation of new records (empty repo_id) and update of existing records (UUID in repo_id). Columns with partner_editable = False are silently skipped during import. A per-row result report is returned after processing. | Encoding: UTF-8 with optional BOM. Partners cannot import or modify records belonging to other partners. |
| *CSV export* | Full export of all own records to a UTF-8 CSV file, containing repo_id and all active columns visible to the partner. The exported file is re-importable. | The export action is logged in the audit trail. |
| *Query resolution* | Partners can view and respond to queries raised by the curation team against specific records. Queries are surfaced within the portal alongside the affected record. | — |

### Table C.2 — Column Visibility and Editability from the Partner Perspective

| partner_visible | partner_editable | Behaviour in the portal |
|---|---|---|
| True | True | Column is visible and editable in the form, inline editor, and CSV import |
| True | False | Column is visible and displayed in the table and form, but the field is locked and cannot be modified |
| False | — | Column is completely absent from the interface; the field is not returned to the client |

### Table C.3 — Actions Logged in the Audit Trail

All partner actions are captured synchronously in the audit log (ActionLog). The log is accessible to curation team administrators only.

| Action | Logged information |
|---|---|
| Record created | Actor, repo_id, field values at creation |
| Record updated (form) | Actor, repo_id, field name, old value, new value |
| Record updated (inline) | Actor, repo_id, field name, old value, new value |
| CSV import completed | Actor, counts of created / updated / skipped records, error list |
| CSV export executed | Actor, number of rows exported |

## D. CRM Data Model

### Figure D.1 — Simplified Entity-Relationship Diagram of the Internal CRM

*High-level entity-relationship diagram of the data model underlying the repository management CRM described in Section 7.2. Five principal entities are shown: the repository record, the metadata schema definition, the legal entity registry, the user profile, and the audit log.*

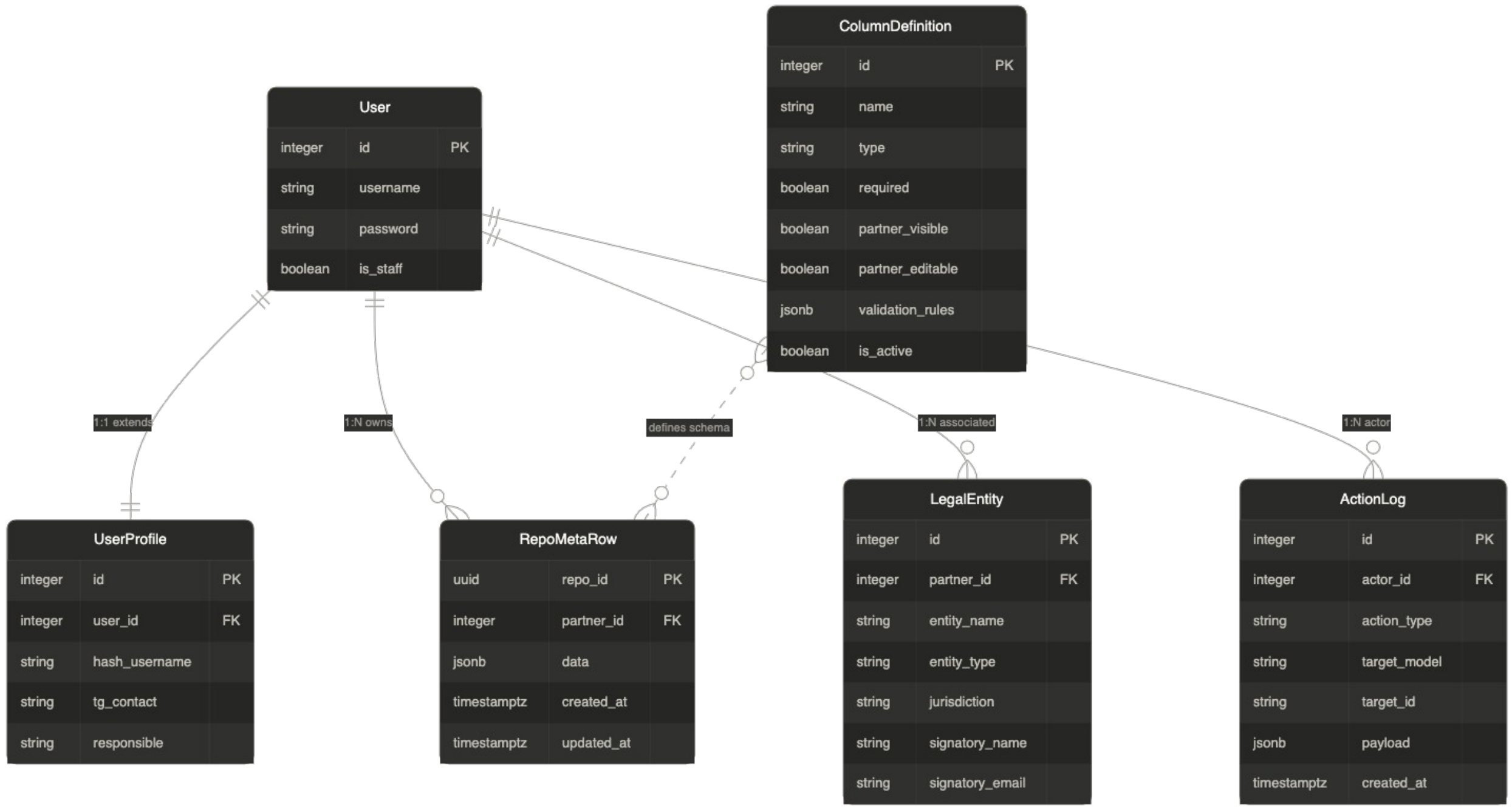

## E. Repository Cloning Script

*Shell script provided to partners for creating Git bundle archives (Section 4.2). The script performs a complete bare-mirror clone of a remote repository and packages it into a* .bundle *file using* git bundle create --all*, capturing all branches, tags, and refs.*

```bash
#!/usr/bin/env bash
set -euo pipefail

INPUT_FILE="${1:-repos.txt}"
MIRRORS_DIR="${2:-./mirrors}"
BUNDLES_DIR="${3:-./partner/bundles}"
OK_FILE="${4:-./gitlab_repos.txt}"

mkdir -p "$MIRRORS_DIR" "$BUNDLES_DIR"
: > "$OK_FILE"

MIRRORS_DIR="$(cd "$MIRRORS_DIR" && pwd)"
BUNDLES_DIR="$(cd "$BUNDLES_DIR" && pwd)"
OK_FILE="$(cd "$(dirname "$OK_FILE")" && pwd)/$(basename "$OK_FILE")"

safe_name() {
 # Produces GitLab-safe and filesystem-safe name WITHOUT underscores.
 # Examples:
 #   https://github.com/org/repo.git          -> org--repo
```

```
 #   git@github.com:org/repo.git              -> org--repo
 #   https://gitlab.com/group/sub/repo         -> group--sub--repo
 local url="$1"
 local s path

 url="${url%.git}"

 s="$(echo "$url" | sed -E 's#^[a-zA-Z]+://##; s#:#/#; s#^[^@]+@##')"
 path="${s#*/}"
 path="${path#/}"

 path="$(echo "$path" | sed -E 's#/+#--#g')"
 path="$(echo "$path" | sed -E 's#[^A-Za-z0-9.-]+#-#g')"
 path="$(echo "$path" | sed -E 's#-+#-#g')"
 path="$(echo "$path" | sed -E 's#^[.-]+##; s#[.-]+$##')"

 if [[ -z "$path" ]]; then
   path="repo"
 fi

 if [[ "$path" == *.git || "$path" == *.atom ]]; then
   path="${path}-repo"
 fi

 echo "$path"
}

repo_only_name() {
 # Extract ONLY the last path segment ("repo") and make it filesystem-safe.
 # Examples:
 #   https://github.com/org/repo.git          -> repo
 #   git@github.com:org/repo.git              -> repo
 #   https://gitlab.com/group/sub/repo        -> repo
 local url="$1"
 local s path base

 # trim spaces
 url="$(echo "$url" | sed -E 's/^[[:space:]]+//; s/[[:space:]]+$//')"

 # drop trailing "/" and trailing ".git"
 url="${url%/}"
 url="${url%.git}"

 # normalize (remove scheme, convert ":" -> "/", drop userinfo)
 s="$(echo "$url" | sed -E 's#^[a-zA-Z]+://##; s#:#/#; s#^[^@]+@##')"

 # drop host
```

```
  path="${s#*/}"
  path="${path#/}"

  # take last segment
  base="${path##*/}"

  # sanitize
  base="$(echo "$base" | sed -E 's#[^A-Za-z0-9.-]+#-#g')"
  base="$(echo "$base" | sed -E 's#-+#-#g')"
  base="$(echo "$base" | sed -E 's#^[.-]+##; s#[.-]+$##')"

  if [[ -z "$base" ]]; then
    base="repo"
  fi

  if [[ "$base" == *.git || "$base" == *.atom ]]; then
    base="${base}-repo"
  fi

  echo "$base"
}

while IFS= read -r repo || [[ -n "$repo" ]]; do
  [[ -z "${repo// }" ]] && continue
  [[ "$repo" =~ ^# ]] && continue

  mirror_name="$(safe_name "$repo")"
  bundle_name="$(repo_only_name "$repo")"

  repo_dir="$MIRRORS_DIR/$mirror_name.git"
  bundle_path="$BUNDLES_DIR/$bundle_name.bundle"

  echo "=== $mirror_name ==="
  echo "repo: $repo"
  echo "bundle: $(basename "$bundle_path")"

  if [[ ! -d "$repo_dir" ]]; then
    echo "→ init bare mirror: $repo_dir"
    git init --bare "$repo_dir" >/dev/null
    git -C "$repo_dir" remote add origin "$repo"
  else
    echo "→ mirror exists, updating"
    current_url="$(git -C "$repo_dir" remote get-url origin 2>/dev/null || true)"
    if [[ "$current_url" != "$repo" && -n "$repo" ]]; then
      echo "→ refresh remote URL (was: ${current_url:-<none>})"
      git -C "$repo_dir" remote set-url origin "$repo"
```

```
  fi
fi

# fetch EVERYTHING the remote advertises
git -C "$repo_dir" config remote.origin.mirror true || true
git -C "$repo_dir" config --unset-all remote.origin.fetch >/dev/null 2>&1 || true
git -C "$repo_dir" config --add remote.origin.fetch "+refs/*:refs/*"
git -C "$repo_dir" config --add remote.origin.fetch "+refs/merge-requests/*:refs/merge-requests/*" || true
git -C "$repo_dir" config --add remote.origin.fetch "+refs/pull/*:refs/pull/*" || true

echo "→ fetch all refs (force/prune)"
git -C "$repo_dir" fetch --force --prune --prune-tags origin

# Ensure bundle has a valid HEAD
remote_head_ref="$(git -C "$repo_dir" ls-remote --symref origin HEAD 2>/dev/null | awk '/^ref:/ {print $2; exit}')"
if [[ -n "$remote_head_ref" ]]; then
  git -C "$repo_dir" symbolic-ref HEAD "$remote_head_ref" || true
else
  fallback_head="$(git -C "$repo_dir" for-each-ref --format='%(refname)' refs/heads | head -n1)"
  [[ -n "$fallback_head" ]] && git -C "$repo_dir" symbolic-ref HEAD "$fallback_head" || true
fi

if ! git -C "$repo_dir" show-ref --quiet; then
  echo "⚠️ no refs fetched (empty repo or no access). skip bundle."
  echo
  continue
fi

echo "→ create bundle: $bundle_path"
rm -f "$bundle_path"
git -C "$repo_dir" bundle create "$bundle_path" --all

echo "$repo" >> "$OK_FILE"

echo "✅ done"
echo
done < "$INPUT_FILE"

echo "OK list: $OK_FILE"
```

## F. Benchmark Decontamination Audit

*The training corpus of the case study (Section 10) was decontaminated by token-level 13-gram overlap — the standard procedure in code-model pre-training (Brown et al., 2020; Li et al.,*

*2023) — against a superset of the benchmarks used in this study, additionally covering MBPP-derived tasks and Rust translations of both suites. The index covers MultiPL-E (HumanEval- and MBPP-derived tasks in PHP, JavaScript, TypeScript, and Rust) and the EvalPlus HumanEval+ and MBPP+ task sets (Python); any file sharing at least one 13-token window with any indexed benchmark solution was removed. In total 907 files were removed from the 404,720 unique files retained after exact deduplication (0.22%): 903 from repositories in the training split and 4 from held-out repositories.*

*Removals concentrate in 10 distinct benchmark tasks. By family: MultiPL-E MBPP translations, 697 files (of which one task, a generic list-partition idiom in its PHP translation, accounts for 380); MultiPL-E HumanEval translations, 153; HumanEval+ (Python), 31; MBPP+ (Python), 26. The dominance of short generic idioms among the matches indicates that most removals are conservative near-matches rather than copied benchmark solutions; we removed them regardless, preferring recall over precision in decontamination. The full removal log (repository, file path, matched benchmark task) is available to licensees together with the dataset.*

## G. Case Study Configuration and Full Results

### G.1 Adaptation configuration

*Base model Qwen2.5-Coder-3B (base variant). Quantization: 4-bit NF4 with double quantization, float16 compute. LoRA: rank 32, α = 32, dropout 0.05, applied to all attention and MLP linear projections (59.9 M trainable parameters, 1.9%). Sequence length 1,024 with document packing; full-sequence next-token loss. Optimizer: paged 8-bit AdamW; learning rate $2\times10^{-4}$, cosine decay, 3% warm-up; weight decay 0; gradient-norm clip 1.0; batch 1 with 16-step gradient accumulation (16,384 tokens per step); 18,127 steps (297 M training tokens; 1.06 epochs over the tokenized, packed corpus — the 296 M figure of Section 10.2 is a byte-based pre-tokenization estimate); float16 with dynamic loss scaling; gradient checkpointing; seed 42. Hardware: one NVIDIA RTX 2080 Ti (11 GB); peak training memory ≈ 6 GB (pre-flight measurement); ≈ 93 h of training compute (estimated from the measured throughput of 895 tokens per second). Checkpoint selection: minimum held-out loss, evaluated every 500 steps on a fixed sample of 399 held-out files (selected: step 17,000).*

### G.2 Evaluation configuration

*Identical for base and adapted models. Generation: greedy decoding, 512 new tokens maximum, completions truncated at function boundaries, seed 42. Functional correctness: MultiPL-E HumanEval-translations for PHP/JavaScript/TypeScript executed in the BigCode evaluation-harness container; Python scored natively with EvalPlus (HumanEval and HumanEval+). Perplexity: per-language exponentiated mean token-level cross-entropy over 1,024-token windows on the held-out split (9,363 files in the four evaluated languages, of 13,827 held-out files across 52 repositories). Completion: 300 examples per language for PHP, JavaScript, TypeScript, Python; for each sampled file, one interior code line is masked (FIM infilling) or*

*used as the continuation target (next line); metrics are exact match and character-level edit similarity; identical example sets for both models.*

### G.3 Full result grids

### Table G.1 — Completion, all languages, tasks, and metrics (base → adapted).

| Language | Task | Exact match | Edit similarity |
|---|---|---|---|
| *PHP* | *FIM infill* | *0.463 → 0.453* | *0.748 → 0.763* |
| *PHP* | *Next line* | *0.333 → 0.353* | *0.664 → 0.684* |
| *JavaScript* | *FIM infill* | *0.570 → 0.563* | *0.819 → 0.803* |
| *JavaScript* | *Next line* | *0.457 → 0.440* | *0.731 → 0.726* |
| *TypeScript* | *FIM infill* | *0.460 → 0.497* | *0.790 → 0.818* |
| *TypeScript* | *Next line* | *0.273 → 0.310* | *0.669 → 0.701* |
| *Python* | *FIM infill* | *0.510 → 0.550* | *0.770 → 0.795* |
| *Python* | *Next line* | *0.420 → 0.403* | *0.713 → 0.711* |

### Table G.2 — Perplexity on held-out repositories (base → adapted).

| Language | Files | Tokens (M) | Base | Adapted | Δ |
|---|---|---|---|---|---|
| *PHP* | *3,658* | *1.68* | *2.122* | *2.010* | *−5.3%* |
| *JavaScript* | *1,685* | *1.59* | *2.003* | *2.052* | *+2.4%* |
| *TypeScript* | *3,170* | *0.92* | *2.573* | *2.356* | *−8.4%* |
| *Python* | *850* | *0.67* | *2.240* | *2.334* | *+4.2%* |

### Table G.3 — Functional correctness with intervals.

*Wilson 95% confidence intervals, reported per model. The MultiPL-E rows use an unpaired two-proportion test (per-task outcomes were not retained in those runs); the Python rows use McNemar's exact test on paired per-task records. Differences are computed from unrounded values.*

| Benchmark | n | Base [95% CI] | Adapted [95% CI] | Test |
|---|---|---|---|---|
| *MultiPL-E PHP* | *161* | *0.497 [0.421, 0.573]* | *0.404 [0.331, 0.481]* | *two-proportion, p ≈ 0.09* |
| *MultiPL-E* | *161* | *0.522 [0.445,* | *0.453 [0.378,* | *two-proportion, p ≈ 0.22* |
| *JavaScript* | | *0.597]* | *0.530]* | |
| *MultiPL-E TypeScript* | *159* | *0.541 [0.463, 0.616]* | *0.409 [0.335, 0.486]* | *two-proportion, p ≈ 0.018* |
| *HumanEval* | *164* | *0.482* | *0.415* | *McNemar exact (21/10 discordant), p = 0.071* |
| *HumanEval+* | *164* | *0.396* | *0.360* | *McNemar exact (17/11 discordant), p = 0.345* |

*Training-dynamics artifacts (training-loss curve, held-out-loss curve) are provided as* analysis/tables/train_loss_curve.csv *and* analysis/tables/heldout_loss_curve.csv*; the source code of the corpus-construction, training, and evaluation pipeline is available to licensees.*

## H. Metric Computation Reference

*This appendix documents the precise computation of every catalog metric. The utility combines five mechanisms, chosen to keep every metric reproducible by rule rather than by judgment:* scc *for line and character accounting (a language-aware code counter with deterministic JSON output, substantially faster than* cloc *on large repositories, which permits re-counting the full catalog on every release);* jscpd *for duplication (a token-based copy-paste detector, so that duplication reflects copied blocks rather than incidental line repetition);* ***Tree-sitter*** *for structural metrics (language-agnostic concrete-syntax parsing across the catalog's language range, via per-language grammars);* ***git*** *itself for history metrics (computed from the repository's own object store, with no external service required); and the* ***hosting-platform APIs*** *(GitHub GraphQL, GitLab REST) for pull-request enrichment where a platform URL is available. Engineering-practice attributes are assigned by rule-based detectors over the repository tree and history; no metric is manually annotated.*

### Table H.1 — Per-metric computation.

| Metric | Mechanism | Computation | Exclusions / thresholds |
|---|---|---|---|
| Raw LoC | scc --format json --no-complexity | Sum of the scc *Lines* column over all source files (physical lines, including blank lines and comments) | none |
| Logical LoC | scc --format json --no-complexity --exclude-dir node_modules,vendor,dist,build,bower_components | Sum of the scc *Code* column (excludes blank and comment lines) | dependency/build directories excluded |
| Auto-generated LoC | scc per-file (--by-file) + generation heuristics | Sum of scc *Code* lines over files classified as generated; never exceeds Logical LoC | classification rules in H.2 |
| | | by construction (the scan is restricted to the Logical-LoC file set) | |
| Language composition | scc per-language totals | Language → fraction of *Code* lines, from scc's language taxonomy | same exclusions as Logical LoC; shares < 1% within a repository omitted; in the released catalog the retained shares sum to one per repository |
| Primary language | derived | Language with the largest share in the composition | — |
| Characters | scc --by-file file set + file read | Sum of Unicode character counts over the Logical-LoC file set | same exclusions as Logical LoC |
| Source file count | scc | scc *Files* column total | none |
| Duplication rate | jscpd --min-tokens 50 --min-lines 5 --max-size 1024mb --reporters json --silent --ignore <H.2 list> | jscpd total duplication percentage / 100, clamped to [0, 1] | generated, minified, dependency, and lock files ignored (list in H.2); 720 s timeout per repository |
| Commit volume | git log --no-merges --oneline | Count of non-merge commits on the default branch, excluding commits whose subject contains "revert" (case-insensitive) | merge and revert commits excluded |
| Total PRs | hosting API, else git fallback | Merged pull/merge requests from the platform API (GitHub GraphQL / GitLab REST) when a platform URL is available and returns a non-zero count; otherwise merge-pattern detection over git log --all (patterns in H.2) | API preferred; fallback deduplicates PR numbers |
| CI on PRs | file presence | "Yes" if any CI configuration path exists (list in H.2) | — |
| Deployment tier | file presence + CI content | Enterprise if IaC/Kubernetes artifacts exist (*.tf, Chart.yaml, | ladder evaluated top-down |
| | | deployment.yaml, k8s/, kubernetes/, *.k8s.y[a]ml); else Full CI-CD if the CI configuration contains a deployment keyword (deploy, release, publish, ship); else Basic CI if CI configuration exists; else None | |
| Observability level | regex scan of source files | Full SRE if a distributed-tracing tool is referenced (OpenTelemetry, Jaeger, Honeycomb); else APM+Alerting if any APM pattern matches (H.2); else Basic if a logger-instantiation pattern matches (H.2); else None | source-file extensions only; vendored paths skipped — tooling referenced only in CI configuration or manifests does not count |
| Test coverage | filename patterns + content verification | Comprehensive if ≥ 10 test files or ≥ 3 distinct test directories; Basic if ≥ 1 test file, or none but a test-framework configuration file exists; else None | compiled-language test files (.cs, .java) must contain framework markers (H.2), not merely a Test-suffixed name |
| Containerized | file presence | "Yes" if any of: Dockerfile, Compose files, or .dockerignore at the root; *.k8s.y[a]ml or Chart.yaml anywhere in the tree; YAML files under deploy/, infra/, k8s/, kubernetes/, or docker/; or a Dockerfile at any depth | — |
| Documentation density | Tree-sitter parse | Fraction of function and method definition nodes | per-language function node types from the grammar map; languages without a |
| | | carrying a leading docstring or an immediately preceding documentation comment | configured grammar are skipped |
| README quality | root README analysis | Comprehensive if (docs/ directory or CONTRIBUTING.md) and README covers setup and usage; Detailed if README covers setup and (usage or architecture); Basic if README > 200 characters; else None (absent, under 50 characters, or 50–200 characters without the keyword sections) | section detection by keyword lists (install/setup/getting started…; usage/example…; architecture/overview/design…) |
| Issue tracking | git log + file presence | Linked to Commits if issue references match in the last 200 non-merge commit subjects across all branches (regex in H.2); upgraded to Full+Design Docs if docs/rfcs, docs/adr, rfcs/, or adr/ exists; Basic if GitHub issue templates exist; else None | — |
| Average function length | Tree-sitter parse | Mean length in lines of the full function-definition node, signature included, over all matched function nodes | same grammar map as documentation density |
| Fork indicator | git configuration + history | 1.0 if an upstream remote is configured, refs/remotes/upstream/* exist, or a merge-commit subject among the last 200 merges matches an upstream-pull pattern (external- | — |

host merge); else 0.0

### H.2 Detector rule sets

***Auto-generation classification.*** *A file is classified as generated if any of: (a) a path component (other than the filename) is one of* generated, __generated__, migrations, .next, .nuxt, out*; (b) the filename matches* *_generated.*, *_pb2.py, *.pb.go, *.min.js, *.min.css, *or* *.bundle.js *(case-insensitive); (c) the filename is exactly* package-lock.json, yarn.lock, Cargo.lock, *or* go.sum *(matched case-insensitively); (d) one of the markers "code generated by" or "do not edit" occurs within the first five lines of the file.*

jscpd ***ignore list.*** *Directory patterns* **/vendor/**, **/node_modules/**, **/dist/**, **/build/**, **/__generated__/**, **/migrations/**, **/generated/***; filename patterns* **/*_generated.*, **/*_pb2.py, **/*.pb.go, **/*.min.js, **/*.min.css, **/*.bundle.js*; lock files* package-lock.json, yarn.lock, Cargo.lock, go.sum, poetry.lock, pnpm-lock.yaml. *Rationale: duplication should measure copied hand-written code, not generated or vendored artifacts.*

***CI configuration paths.*** .github/workflows, .circleci, .travis.yml, Jenkinsfile, .gitlab-ci.yml, azure-pipelines.yml, .appveyor.yml, .drone.yml, bitbucket-pipelines.yml, .buildkite, circle.yml.

***APM patterns (observability).*** *Word-boundary regular expressions for: sentry / sentry_sdk, datadog / ddtrace, newrelic / new_relic, prometheus_client, pagerduty, opsgenie, honeycomb, jaeger, opentelemetry / @opentelemetry/, elastic_apm, rollbar, bugsnag, raygun, instana, dynatrace; raygun is matched as* raygun4* *or* raygun.io. *The Full SRE tier is triggered by the distributed-tracing subset {opentelemetry, jaeger, honeycomb}. Basic-tier logging patterns require logger instantiation:* console.log( / console.error( / console.warn( *(JavaScript),* logging.basicConfig / logging.getLogger *(Python),* logrus.New / zap.New* *(Go),* winston.createLogger / bunyan.createLogger / pino( *(Node.js),* log4j.getLogger / Logger.getLogger */ logback (Java). Matching is restricted to the nine source-file extensions .py, .js, .ts, .go, .java, .cs, .rb, .php, .rs and skips vendored directories, so that a tool mentioned only in CI configuration or a dependency manifest is not counted as adopted.*

***Test-file patterns.*** test_*.py, *_test.py, *_test.go, *.spec.ts, *.spec.js, *.test.ts, *.test.js, *Test.java, *Spec.java, *_test.rb, *_spec.rb, *_test.rs, *_test.cs, *Test.cs. *For .cs and .java the file content must additionally contain a test framework marker (NUnit / MSTest / xUnit / Unity Test Framework attributes or* using/import *statements; JUnit / TestNG / AndroidX annotations or imports), preventing production classes with Test-suffixed names from counting. Test-framework configuration files accepted for the Basic tier:* pytest.ini, jest.config.{js,ts,mjs}, .mocharc.{js,yml}, karma.conf.js, phpunit.xml[.dist], rspec, .rspec.

***Issue-reference pattern.*** *Case-insensitive regular expression over the last 200 non-merge commit subjects:* (fixes?|closes?|resolves?) #N, *bare* #N, JIRA-*, LINEAR-*, *or a generic ticket key* [A-Z]{2,10}-N.

***Pull-request fallback patterns (git history).*** *GitHub merge commits ("Merge pull request #N"), GitHub squash merges (subject ending in "(#N)"), GitLab merge commits ("See merge request*

*…!N"); PR/MR numbers are deduplicated before counting. The reviewed-PR count is populated only from the platform API, as review metadata is absent from git history.*

***Tree-sitter grammar map.*** *Grammars are loaded per file extension from a configured extension→language map (60+ extensions) with per-language function node types (for example* function_definition *for C/C++/Python-family grammars;* method_declaration*,* constructor_declaration*, and* local_function_statement *for C#). Files whose language has no configured grammar are skipped and do not contribute to documentation density or function length.*

### H.3 Design rationale

*Three properties guided the design. Reproducibility: every metric is a deterministic function of the repository content, so any licensee can re-derive the catalog values from the delivered repositories with the pinned utility version. Conservatism where detection is heuristic: content verification for compiled-language test files, source-only scanning for observability, and the restriction of the auto-generated scan to the logical-LoC file set all bias the attributes toward under- rather than over-claiming engineering maturity. Separation of measurement generations: metrics whose definition changed between utility versions are reported under new names and never mixed with version-1 figures (Section 4.3.2, "Changes relative to the earlier characterization").*